\documentclass[11pt]{article}

\usepackage{tikz,algonjk,graphicx,units,url}
\usepackage{amsthm,amsfonts,amsmath,times}
\usepackage[justification=centering]{caption}

\advance \topmargin by -1.0in
\advance \oddsidemargin by -0.8in
\newcommand{\myparskip}{3pt}
\parskip \myparskip

\newtheorem{lemma}{Lemma}[section]
\newtheorem{theorem}[lemma]{Theorem}

\newtheorem{definition}[lemma]{Definition}

\newtheorem{prop}[lemma]{Proposition}
\newtheorem{claim}[lemma]{Claim}
\newtheorem{remark}[lemma]{Remark}

\newtheorem{conjecture}{Conjecture}

\newtheorem*{redlem}{Reduction Lemma}

\newcommand{\opt}{\text{\sc OPT}}
\newcommand{\sskconn}{{\sc SS-$k$-Connectivity}}
\newcommand{\econn}{\lambda}
\newcommand{\vconn}{\kappa}
\newcommand{\elconn}{\kappa'}
\newcommand{\etal}{{\em et al.}\ }

\renewenvironment{proof}{\vspace{-0.1in}\noindent{\bf Proof:}}%
        {\hspace*{\fill}$\Box$\par}
\newenvironment{proofof}[1]{\smallskip\noindent{\bf Proof of #1:}}%
        {\hspace*{\fill}$\Box$\par}
\newenvironment{proofsketch}{\vspace{-0.1in}\noindent{\bf Proof Sketch:}}%
        {\hspace*{\fill}$\Box$\par}

\def\eps{\varepsilon}

\def\floor#1{\lfloor {#1} \rfloor}
\def\ceil#1{\lceil {#1} \rceil}

\def\script#1{\mathcal{#1}}

\def\wc{\kappa'}

\title{A Graph Reduction Step Preserving Element-Connectivity \\ and Applications}
\author{
Chandra Chekuri\thanks{Dept. of Computer Science, University of Illinois, 
Urbana, IL 61801. Partially supported by NSF grants CCF 07-28782 and 
CNS-0721899. {\tt chekuri@cs.uiuc.edu}}
  \and 
Nitish Korula\thanks{Dept. of Computer Science, University of Illinois, Urbana,
    IL 61801. Partially supported by NSF grant CCF 07-28782. {\tt
      nkorula2@illinois.edu}} 
}

\begin{document}
\maketitle

\begin{abstract}
  Given an undirected graph $G=(V,E)$ and subset of terminals $T
  \subseteq V$, the {\em element-connectivity} $\elconn_G(u,v)$ of two
  terminals $u,v \in T$ is the maximum number of $u$-$v$ paths that
  are pairwise disjoint in both edges and non-terminals $V \setminus
  T$ (the paths need not be disjoint in terminals).
  Element-connectivity is more general than edge-connectivity and less
  general than vertex-connectivity. Hind and Oellermann \cite{hind}
  gave a graph reduction step that preserves the {\em global}
  element-connectivity of the graph.  We show that this step also
  preserves {\em local} connectivity, that is, all the pairwise
  element-connectivities of the terminals. We give two applications of
  this reduction step to connectivity and network design problems.
  \begin{itemize}
  \item Given a graph $G$ and disjoint terminal sets $T_1, T_2,
    \ldots, T_m$, we seek a maximum number of element-disjoint
    Steiner forests where each forest connects each $T_i$. We prove
    that if each $T_i$ is $k$ element connected then there exist
    $\Omega(\frac{k}{\log h \log m})$ element-disjoint Steiner
    forests, where $h = |\bigcup_i T_i|$. If $G$ is planar (or more
    generally, has fixed genus), we show that there exist $\Omega(k)$
    Steiner forests. Our proofs are constructive, giving poly-time
    algorithms to find these forests; these are the first non-trivial
    algorithms for packing element-disjoint Steiner Forests.
  
  \item We give a very short and intuitive proof of a
    spider-decomposition theorem of Chuzhoy and Khanna
    \cite{ChuzhoyK08} in the context of the single-sink $k$-{\em
      vertex-connectivity} problem; this yields a simple and
    alternative analysis of an $O(k \log n)$ approximation.
  \end{itemize}
  Our results highlight the effectiveness of the element-connectivity
  reduction step; we believe it will find more applications in the future.
\end{abstract}


\section{Introduction}
In this paper we consider several connectivity and network design
problems. Given an undirected graph $G$ and two nodes $u,v$ we let
$\econn_G(u,v)$ and $\vconn_G(u,v)$ denote the edge and vertex
connectivities between $u$ and $v$ in $G$. It is well-known that
edge-connectivity problems are ``easier'' than their
vertex-connectivity counterparts. Vertex-connectivity exhibits less
structure than edge-connectivity and this often translates into
significant differences in the algorithmic and computational
difficulty of the corresponding problems. As an example, consider the
well-known survivable network design problem (SNDP): the input
consists of an undirected edge-weighted graph $G$ and connectivity
requirements $r:V \times V \rightarrow Z^+$ between each pair of
vertices.  The goal is to find a min-cost subgraph $H$ of $G$ such
that each pair $u,v$ has $r(u,v)$ disjoint paths between them in
$H$. If the paths are required to be edge-disjoint ($\econn_H(u,v) \ge
r(u,v)$) then the problem is referred to as EC-SNDP and if the paths
are required to be vertex-disjoint the problem is referred to as
VC-SNDP.  Jain \cite{Jain} gave a $2$-approximation for EC-SNDP based
on the powerful iterated rounding technique. On the other hand,
VC-SNDP is known to be hard to within polynomial factors
\cite{KortsarzKL,ChakCK08}.  To address this gap, Jain \etal
\cite{JainMVW99} introduced a connectivity measure intermediate to
edge and vertex connectivities known as {\em element-connectivity}.
The vertices are partitioned into terminals $T \subseteq V$ and
non-terminals $V \setminus T$. The element-connectivity between two
terminals $u,v$, denoted by $\elconn_G(u,v)$ is defined to be the
maximum number of paths between $u$ and $v$ that are pairwise disjoint
in edges and non-terminals (the paths can share terminals). In some
respects, element-connectivity resembles edge-connectivity: For
example, $\elconn(u,w) \ge \min(\elconn(u,v), \elconn(v,w))$ for any
three terminals $u,v,w$; this triangle inequality holds for
edge-connectivity but does not for vertex-connectivity. In
element-connectivity SNDP (ELC-SNDP) the requirements are only between
terminals and the goal is to find a min-cost subgraph $H$ such that
$\elconn_H(u,v) \ge r(u,v)$ for each $u,v \in T$. Fleischer, Jain and
Williamson \cite{FleischerJW} (see also \cite{CheriyanVV06})
generalized the iterated rounding technique of Jain for EC-SNDP to
give a $2$-approximation for ELC-SNDP.  In other respects,
element-connectivity is related to vertex connectivity. One class of
problems motivating this paper is on generalizing the classical
theorem of Menger on $s$-$t$ vertex-connectivity; we discuss this
below.

In studying element-connectivity, we often assume without loss of
generality that there are no edges between terminals (by subdividing
each such edge) and hence $\elconn(u,v)$ is the maximum number of
non-terminal disjoint $u$-$v$ paths. Menger's theorem shows that the
maximum number of internally vertex-disjoint $s$-$t$ paths is equal to
$\vconn(s,t)$.  Hind and Oellermann \cite{hind} considered a natural
generalization to multiple terminals. Given a terminal set $T
\subseteq V$, what is the maximum number of trees that each contain
$T$ and are disjoint in $V \setminus T$? The natural upper bound here
is the element connectivity of $T$ in $G$, in other words, $k =
\min_{u,v \in T} \elconn(u,v)$. In \cite{hind} a graph reduction step
was introduced to answer this question.  Cheriyan and Salavatiour
\cite{cs} called this the problem of packing element-disjoint Steiner
trees; crucially using the graph reduction step, they showed that
there always exist $\Omega(k/ \log |T|)$ element-disjoint Steiner
trees and moreover, this bound is tight (up to constant factors) in
the worst case. In contrast, if we seek edge-disjoint Steiner trees
then Lau \cite{Lau1} has shown that if $T$ is $26k$ edge-connected in
$G$, there are $k$ edge-disjoint trees each of which spans $T$.

Finally, we remark that in some recent work Chuzhoy and Khanna
\cite{ChuzhoyK08} gave an $O(k \log |T|)$ approximation for the
special case of VC-SNDP in which a terminal set $T$ needs to be
$k$-vertex-connected (this is equivalent to the single-sink problem).
Their algorithm and analysis are based on a structural
characterization of feasible solutions --- they use
element-connectivity (they call it weak connectivity) as a key
stepping stone. Subsequent to this paper, Chuzhoy and Khanna
\cite{CKsndp} gave a simple and elegant reduction from the the
\emph{general} VC-SNDP problem to ELC-SNDP, obtaining an $O(k^3 \log
n)$-approximation and reinforcing the connection between element- and
vertex-connectivity.

The discussion above suggests that it is fruitful to study
element-connectivity as a way to generalize edge-connectivity and
attack problems on vertex-connectivity.  In this paper we consider the
graph reduction step for element-connectivity introduced by Hind and
Oellermann \cite{hind} (and rediscovered by Cheriyan and Salavatipour
\cite{cs}). We generalize the applicability of the step and
demonstrate applications to several problems.

\smallskip
\noindent
{\bf A Graph Reduction Step Preserving Element Connectivity:}
The well-known {\em splitting-off} operation introduced by Lov\'{a}sz
\cite{Lovasz} is a standard tool in the study of (primarily)
edge-connectivity problems. Given an undirected multi-graph $G$ and
two edges $su$ and $sv$ incident to $s$, the splitting-off operation
replaces $su$ and $sv$ by the single edge $uv$.  Lov\'{a}sz proved the
following theorem on splitting-off to preserve {\em global}
edge-connectivity.

\begin{theorem}[Lov\'{a}sz]
  Let $G=(V\cup\{s\}, E)$ be an undirected multi-graph in which $V$ is
  $k$-edge-connected for some $k \ge 2$ and degree of $s$ is
  even. Then for every edge $su$ there is another edge $sv$ such that
  $V$ is $k$-edge-connected after splitting-off $su$ and $sv$.
\end{theorem}

Mader strengthened the above theorem to show the existence of
a pair of edges incident to $s$ that when split-off preserve the
{\em local} edge-connectivity of the graph. 

\begin{theorem}[Mader \cite{Mader}]
  \label{thm:mader}
  Let $G=(V\cup \{s\},E)$ be an undirected multi-graph, where $deg(s)
  \neq 3$ and $s$ is not incident to a cut edge of $G$.  Then $s$
  has two neighbours $u$ and $v$ such that the graph $G'$ obtained
  from $G$ by replacing $su$ and $sv$ by $uv$ satisfies
  $\lambda_{G'}(x,y) = \lambda_G(x,y)$ for all $x,y \in V \setminus
  \{s\}$.
\end{theorem}

Generalization to directed graphs are also known
\cite{Mader,Frank,Jackson}. The splitting-off theorems have numerous
applications in graph theory and combinatorial optimization. See
\cite{Lovasz,Frank92,Kriesell,JainMS03,ChekuriS,Lau1,Lau2,LauK} for
various pointers and applications.  Although splitting-off techniques
can be sometimes be used in the study of vertex-connectivity, their
use is limited and no generally applicable theorem akin to
Theorem~\ref{thm:mader} is known. On the other hand, Hind and
Oellermann \cite{hind} proved an elegant theorem on preserving global
element connectivity. In the sequel we use $\elconn_G(S)$ to denote
$\min_{u,v \in S} \elconn_G(u,v)$ and $G/pq$ to denote the graph
obtained from $G$ by contracting vertices $p,q$.

\begin{theorem}[Hind \& Oellermann \cite{hind}]
  \label{thm:ho}
  Let $G=(V,E)$ be an undirected graph and $T \subseteq V$ be a terminal-set
  such that $\elconn_G(T) \ge k$ for each pair $u,v \in T$. 
  Let $(p,q)$ be {\em any} edge where $p,q \in V \setminus T$. Then
  $\elconn_{G_1}(T) \ge k$ or $\elconn_{G_2}(T) \ge k$ where $G_1 = G - pq$
  and $G_2 = G/pq$. 
\end{theorem}

This theorem has been used in two applications on element-connectivity
\cite{cs,LauK}. We generalize it to handle local connectivity,
increasing its applicability. 

\begin{redlem} \label{lem:reduction}
  Let $G=(V,E)$ be an undirected graph and $T \subseteq V$ be a terminal-set.
  Let $(p,q)$ be {\em any} edge where $p,q \in V \setminus T$ and
  let $G_1 = G - pq$ and $G_2 = G/pq$. Then one of the following holds:
  (i) $\forall u,v \in T$, $\elconn_{G_1}(u,v) = \elconn_{G}(u,v)$ 
  (ii) $\forall u,v \in T$, $\elconn_{G_2}(u,v) = \elconn_{G}(u,v)$.
\end{redlem}

\begin{remark}
  The Reduction Lemma, applied repeatedly, transforms a graph
  into another graph in which the non-terminals form a stable set. Moreover,
  the reduced graph is a minor of the original graph.
\end{remark}

We give applications of the Reduction Lemma (using additional ideas)
to two problems that we had briefly alluded to already. We discuss
these below.

\smallskip
\noindent {\bf Packing Element-Disjoint Steiner Trees and Forests:} 
There has been much interest in the recent past on algorithms for
(integer) packing of disjoint Steiner trees in both the edge and
element-connectivity settings
\cite{Kriesell,JainMS03,Lau1,Lau2,CheriyanS04,cs,ChekuriS}. (A
\emph{Steiner tree} is simply a tree containing the entire terminal
set $T$.) See \cite{GrotschelMW97} for applications of Steiner tree
packing to VLSI design. An outstanding open problem is Kriesell's
conjecture which states that if the terminal set $T$ is
$2k$-edge-connected then there are $k$-edge-disjoint Steiner trees
each of which spans $T$; this would generalize a classical theorem of
Nash-Williams and Tutte on edge-disjoint spanning trees. Lau made
substantial progress \cite{Lau1} and proved that $26k$-connectivity
suffices for $k$ edge-disjoint Steiner trees; he extended his result
for packing Steiner forests \cite{Lau2}.  We remark that Mader's
splitting-off theorem plays an important role in Lau's work. The
element-disjoint Steiner tree packing problem was first considered by
Hind and Oellermann. As we mentioned, Cheriyan and Salavatipour
\cite{cs} gave a nearly tight bound for this problem. Their result relies
crucially on Theorem~\ref{thm:ho} followed by a simple randomized
coloring algorithm whose analysis extends a similar algorithm for
computing the domatic number of a graph \cite{FeigeHKS00}. In
\cite{CalinescuCV07} the random coloring idea was shown to apply more
generally in the context of packing bases of an arbitrary monotone
submodular function; in addition, a derandomization was provided in
\cite{CalinescuCV07} via the use of min-wise independent
permutations. It is also known that the problem of packing
element-disjoint Steiner trees is hard to approximate to within an
$\Omega(\log n)$ factor \cite{CheriyanS04}. Here, we consider the more
general problem of packing Steiner forests that was posed by
\cite{cs}.  The input consists of a graph $G=(V,E)$ and disjoint
terminal sets $T_1, T_2, \ldots, T_m,$ such that $\elconn_G(T_i) \ge
k$ for $1 \le i \le k$.  What is the maximum number of element
disjoint forests such that in each forest $T_i$ is connected for $1
\le i \le k$? Our local connectivity reduction step is primarily
motivated by this question.  For general graphs we prove that there
exist $\Omega(k/(\log |T| \log m))$ element disjoint forests, where $T
= \bigcup_i T_i$. This can also be viewed as an $O(\log |T| \log m)$
approximation for the problem. We apply the Reduction Lemma to obtain
a graph in which the non-terminals are a stable set. We cannot however
apply the random coloring approach directly --- in fact we can show
that it does not work. Instead we decompose the graph into highly
connected subgraphs and then apply the random coloring approach in
each subgraph separately.

We also study the packing problem in planar graphs and graphs of fixed
genus, and prove substantially stronger results. Here too, the first
step is to use the Reduction Lemma (recall that the reduced graph is a
minor of the original graph and hence is also planar). After the
reduction step, we employ a very different approach from the one for
general graphs.  Our main insight is that planarity restricts the
ability of non-terminals to provide high element-connectivity to the
terminals. We formalize this intuition by showing that there are some
two terminals $u,v$ that have $\Omega(k)$ parallel edges between them
which allows us to contract them and recurse. Using these ideas, for
planar graphs we prove that there exist $\ceil{k/5} -1$ disjoint
forests. Our method also extends to give an $\Omega(k)$ bound for
graphs of a fixed genus, and we conjecture that one can find
$\Omega(k)$ disjoint forests in graphs excluding a fixed minor; we
give evidence for this by proving it for packing Steiner trees in
graphs of fixed treewidth. Note that these bounds also imply
corresponding approximation algorithms for maximizing the number of
disjoint forests. These are the first non-trivial bounds for packing
element-disjoint Steiner forests in general graphs or planar
graphs. Since element-connectivity generalizes edge-connectivity, our
bounds in planar graphs are considerably stronger than those of given
by Lau \cite{Lau1,Lau2} for {\em edge}-connectivity. Our proof is
simple, however, we remark that the simplicity of the proof comes from
thinking about element-connectivity (using the Reduction Lemma)
instead of edge-connectivity! Our proof also gives the strong property
that the non-terminals in the forests all have degree $2$.

\smallskip
\noindent {\bf Single-Sink $k$-vertex-connectivity:} Polynomial
factor inapproximability results for VC-SNDP
\cite{KortsarzKL,ChakCK08} have focused attention on restricted, yet
useful, special cases of the problem. In recent work Chakraborty,
Chuzhoy and Khanna \cite{ChakCK08} considered the single-sink
$k$-vertex-connectivity problem for small $k$; the goal is to
$k$-vertex-connect a set of terminals $T$ to a given root $r$. This
problem is approximation-equivalent to the subset $k$-connectivity
problem in which $T$ needs to be $k$-connected \cite{ChakCK08}.  If
$k=1$, this is the NP-Hard Steiner tree problem and a $2$-approximation
is well-known. For $k=2$, a $2$-approximation follows from
\cite{FleischerJW} whose algorithm can handle the more general VC-SNDP
with requirements in $\{0,1,2\}$. For $k > 2$ the first non-trivial
approximation algorithm was given in \cite{ChakCK08}; the
approximation ratio was $k^{O(k^2)} \log^4 n$. Improvements were given
in \cite{ChuzhoyK08,ChekuriK08} with Chuzhoy and Khanna
\cite{ChuzhoyK08} achieving the currently best known approximation
ratio of $O(k \log |T|)$.  The algorithms are essentially the same in
\cite{ChakCK08,ChuzhoyK08,ChekuriK08} and build upon the insights from
\cite{ChakCK08}; the analysis in \cite{ChuzhoyK08} relied on a
beautiful decomposition result for $k$-connectivity which is
independently interesting from a graph theoretic view point. The proof
of this theorem in \cite{ChuzhoyK08} is long and complicated although
it is based on only elementary operations. Using the Reduction Lemma,
we give an alternate proof of the main technical result which is only
half a page long! We mention that the decomposition theorem has
applications to more general network design problems such as the
rent-or-buy and buy-at-bulk network design problems as shown in
\cite{ChekuriK08}.  Due to space constraints we omit these
applications in this paper.

\smallskip
\noindent {\bf Related Work:} We have already mentioned most of the
closely related papers. Our work on packing Steiner forests in planar
graphs was inspired by a question by Joseph Cheriyan
\cite{Cheriyan}. Independent of our work, Aazami, Cheriyan and Jampani
\cite{ACJ08} proved that if a terminal set $T$ is
$k$-element-connected in a planar graph then there exist $k/2 -1$
element-disjoint Steiner trees, and moreover this is tight. They also
prove that it is NP-hard to obtain a $(1/2+\eps)$ approximation for
this problem. Our bound for packing Steiner Trees in planar graphs is
slightly weaker than theirs; however, our algorithms and proofs are
simple and intuitive, and generalize to packing Steiner forests. Their
algorithm uses Theorem~\ref{thm:ho}, followed by a reduction to a
theorem of Frank \etal \cite{FKK} that uses Edmonds' matroid partition
theorem. One could attempt to pack Steiner forests using their
approach (with the stronger Reduction Lemma in place of
Theorem~\ref{thm:ho}), but the theorem of \cite{FKK} does not have a
natural generalization for Steiner forests.  The techniques of both
\cite{ACJ08} and this paper extend to graphs of small genus or
treewidth; we discuss this further in
Section~\ref{subsec:planarPacking}.  We refer the reader to
\cite{ChakCK08,ChuzhoyK08,ChekuriK08} for more discussion of recent
work on single-sink vertex connectivity, including hardness results
\cite{ChakCK08} and extensions to related problems such as the
node-weighted case \cite{ChuzhoyK08} and buy-at-bulk network design
\cite{ChekuriK08}. Nutov \cite{Nutov09} has recently given alternate
algorithms, based on the primal-dual method, for the single-sink
vertex-connectivity network design with approximation ratios
comparable to those from \cite{ChuzhoyK08}. These algorithms do not
have have the advantage of the structural decomposition of
\cite{ChuzhoyK08}. We mention that if $T=V$, that is, we wish to find
a min-cost subgraph of $G$ that is $k$-connected then an $O(\log^2 k)$
approximation is known \cite{FakL08,KortsarzN05,CheriyanVV03}. We also
refer the reader to a survey on network design by Kortsarz and Nutov
\cite{KortsarzN_survey}.

\section{The Reduction Lemma}
Let $G(V,E)$ be a graph, with a given set $T \subseteq V(G)$ of
terminals. For ease of notation, we subsequently refer to terminals as
\emph{black} vertices, and non-terminals (also called Steiner
vertices) as \emph{white}. The elements of $G$ are white vertices and
edges; two paths are \emph{element-disjoint} if they have no white
vertices or edges in common. Recall that the element-connectivity of
two black vertices $u$ and $v$, denoted by $\elconn_G(u,v)$, is the
maximum number of element-disjoint (that is, disjoint in edges and
white vertices) paths between $u$ and $v$ in $G$.  We omit the
subscript $G$ when it is clear from the context.

For this section, to simplify the proof, we will assume that $G$ has
no edges between black vertices; any such edge can be subdivided, with
a white vertex inserted between the two black vertices. It is easy to
see that two paths are element-disjoint in the original graph iff they
are element-disjoint in the modified graph. Thus, we can say that
paths are element disjoint if they share no white vertices, or that
$u$ and $v$ are $k$-element-connected if the smallest set of white
vertices whose deletion separates $u$ from $v$ has size $k$.

Recall that our lemma generalizes Theorem~\ref{thm:ho} on preserving
global connectivity. We remark that our proof is based on a cutset
argument unlike the path-based proofs in \cite{hind,cs} for the global
case.

\begin{redlem}
  Given $G(V,E)$ and $T$, let $pq \in E(G)$ be any edge such that $p$
  and $q$ are both white. Let $G_1 = G - pq$ and $G_2 = G / pq$ be the
  graphs formed from $G$ by deleting and contracting $pq$
  respectively. Then, 
  (i) $\forall u,v \in T, \wc_{G_1}(u,v) = \wc_G(u,v)$ or 
  (ii) $\forall u,v \in T, \wc_{G_2}(u,v) = \wc_G(u,v)$.
\end{redlem}
\vspace{0.1in}
\begin{proof}
  Consider an arbitrary edge $pq$. Deleting or contracting an edge can
  reduce the element-connectivity of a pair by at most $1$.  Suppose
  the lemma were not true; there must be pairs $s,t$ and $x,y$ of
  black vertices such that $\elconn_{G_1}(s,t) = \elconn_G(s,t) - 1$
  and $\elconn_{G_2}(x,y) = \elconn_G(x,y) - 1$. The pairs have to be
  distinct since it cannot be the case that $\elconn_{G_1}(u,v) =
  \elconn_{G_2}(u,v) = \elconn_G(u,v) - 1$ for any pair $u,v$. (To see
  this, if one of the $\elconn_G(u,v)$ $u$-$v$ paths uses $pq$,
  contracting the edge will not affect that path, and will leave the
  other paths untouched. Otherwise, no path uses $pq$, and so it can
  be deleted.). Note that one of $s,t$ could be the same vertex as one
  of $x,y$; for simplicity we will assume that $\{s,t\} \cap \{x,y\} =
  \emptyset$, but this does not change our proof in any detail.  We
  show that our assumption on the existence of $s,t$ and $x,y$ with
  the above properties leads to a contradiction. Let $\elconn_{G}(s,t)
  = k_1$ and $\elconn_G(x,y) = k_2$. We use the following facts
  several times. 
  
  \vspace{-0.15in}
  \begin{enumerate}
    \item Any cutset of size less than $k_1$ that separates $s$ and
      $t$ in $G_1$ cannot include $p$ or $q$. (If it did, it would
      also separate $s$ and $t$ in $G$.)

    \item $\elconn_{G_1}(x,y) = k_2$ since $\elconn_{G_2}(x,y) = k_2 - 1$.
  \end{enumerate}

  We define a vertex tri-partition of a graph $G$ as follows:
  $(A,B,C)$ is a vertex tri-partition of $G$ if $A,B,$ and $C$
  partition $V(G)$, $B$ contains only white vertices, and there are no
  edges between $A$ and $C$. (That is, removing the white vertices in
  $B$ disconnects $A$ and $C$.)

  Since $\elconn_{G_1}(s,t) = k_1 - 1$, there is a
  vertex-tri-partition $(S,M,T)$ such that $|M| = k_1-1$ and $s \in S$
  and $t \in T$.  From Fact 1 above, $M$ cannot contain $p$ or
  $q$. For the same reason, it is also easy to see that $p$ and $q$
  cannot be both in $S$ (or both in $T$); otherwise $M$ would be a
  cutset of size $k_1-1$ in $G$. Therefore, assume w.l.o.g. that $p
  \in S, q \in T$.
  
  Similarly, since $\elconn_{G_2}(x,y) = k_2 - 1$, there is a
  vertex-tri-partition $(X,N',Y)$ in $G_2$ with $|N'| = k_2 - 1$ and
  $x \in X$ and $y \in Y$. We claim that $N'$ contains the contracted
  vertex $pq$ for otherwise $N'$ would be a cutset of size $k_2-1$ in
  $G$. Therefore, it follows that $(X,N,Y)$ where $N = N' \cup
  \{p,q\}- \{pq\}$ is a vertex-tri-partition in $G$ that separates $x$
  from $y$. Note that $|N| = k_2$ and $N$ includes {\em both} $p$ and $q$.
  For the latter reason we note that $(X,N,Y)$ is a vertex-tri-partition
  also in $G_1$.

  Subsequently, we work with the two vertex tri-partitions $(S,M,T)$
  and $(X,N,Y)$ in $G_1$ (we stress that we work in $G_1$ and not in
  $G$ or $G_2$). Recall that $s,p \in S$, and $t, q \in T$, and that
  $M$ has size $k_1 - 1$; also, $N$ separates $x$ from $y$, and $p, q
  \in N$.  Fig. 1 (a) below shows these vertex tri-partitions. Since
  $M$ and $N$ contain only white vertices, all terminals are in $S$ or
  $T$, and in $X$ or $Y$. We say that $S \cap X$ is \emph{diagonally
    opposite} from $T \cap Y$, and $S \cap Y$ is diagonally opposite
  from $T \cap X$. Let $A,B,C,D$ denote $S \cap N, X \cap M, T \cap N$
  and $Y \cap M$ respectively, with $I$ denoting $N \cap M$; note that
  $A,B,C,D,I$ partition $M \cup N$.

  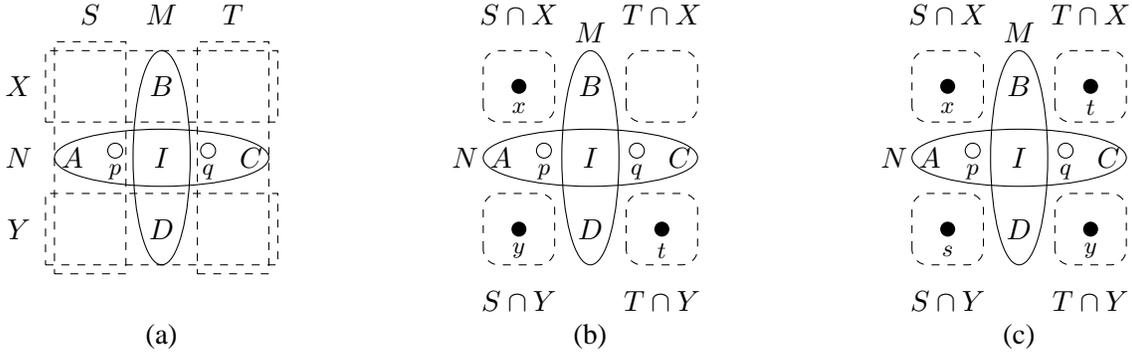
\begin{figure}[h]
    \centering
    \begin{center}
      \begin{tikzpicture}[scale=0.95]

        \tikzstyle{dot}=[circle,inner sep=2pt,fill=black];
        \tikzstyle{elem}=[circle,draw,inner sep=2pt];
        
        \begin{scope}[xshift=-6cm] 
          \draw[dashed] (0,-0.125) rectangle (1,3.125);
          \draw (1.5,1.5) ellipse (0.4cm and 1.5cm);
          \draw[dashed] (2,-0.125) rectangle (3,3.125);
          \node at (0.5,3.5) {$S$}; \node at (1.5,3.5) {$M$}; \node at
          (2.5,3.5) {$T$};

          \draw[dashed] (-0.125,0) rectangle (3.125,1);
          \draw (1.5,1.5) ellipse (1.5cm and 0.4cm);
          \draw[dashed] (-0.125,2) rectangle (3.125,3);
          \node at (-0.5,2.5) {$X$}; \node at (-0.5,1.5) {$N$}; \node at
          (-0.5,0.5) {$Y$};

          \node at (0.25,1.5) {$A$}; \node at (1.5,2.5) {$B$}; \node at
          (2.75,1.5) {$C$}; \node at (1.5,0.5) {$D$}; \node at (1.5,1.5)
          {$I$};

          \node at (0.85,1.6) [elem] {}; \node at (0.85,1.3) [font=\footnotesize]{$p$};
          \node at (2.15,1.6) [elem] {}; \node at (2.15,1.3) [font=\footnotesize]{$q$};

          \node at (1.5,-1) {(a)};
        \end{scope}

        \begin{scope}[xshift=0cm] 
          \draw (1.5,1.5) ellipse (0.4cm and 1.5cm); \node at (-0.25,1.5) {$N$};
          \draw (1.5,1.5) ellipse (1.5cm and 0.4cm); \node at (1.5,3.25) {$M$};
          \node at (0.25,1.5) {$A$}; \node at (1.5,2.5) {$B$};
          \node at (2.75,1.5) {$C$}; \node at (1.5,0.5) {$D$};
          \node at (1.5,1.5) {$I$};

          \node at (0.85,1.6) [elem] {}; \node at (0.85,1.3) [font=\footnotesize]{$p$};
          \node at (2.15,1.6) [elem] {}; \node at (2.15,1.3) [font=\footnotesize]{$q$};

          \draw [dashed,rounded corners=6pt] (0,0) rectangle (1,1);
          \draw [dashed,rounded corners=6pt] (2,0) rectangle (3,1);
          \draw [dashed,rounded corners=6pt] (0,2) rectangle (1,3);
          \draw [dashed,rounded corners=6pt] (2,2) rectangle (3,3);
          
          \node at (0.5,3.5) {$S \cap X$}; \node at (2.5,3.5) {$T \cap X$};
          \node at (0.5,-0.5) {$S \cap Y$}; \node at (2.5,-0.5) {$T \cap Y$};

          \node [dot] at (0.5,2.5) {}; \node at (0.5,2.2) [font=\footnotesize]{$x$};
          \node [dot] at (0.5,0.5) {}; \node at (0.5,0.2) [font=\footnotesize]{$y$};
          \node [dot] at (2.5,0.5) {}; \node at (2.5,0.2) [font=\footnotesize]{$t$};

          \node at (1.5,-1) {(b)};
          
        \end{scope}

        \begin{scope}[xshift=6cm] 
          \draw (1.5,1.5) ellipse (0.4cm and 1.5cm); \node at (-0.25,1.5) {$N$};
          \draw (1.5,1.5) ellipse (1.5cm and 0.4cm); \node at (1.5,3.25) {$M$};
          \node at (0.25,1.5) {$A$}; \node at (1.5,2.5) {$B$};
          \node at (2.75,1.5) {$C$}; \node at (1.5,0.5) {$D$};
          \node at (1.5,1.5) {$I$};

          \node at (0.85,1.6) [elem] {}; \node at (0.85,1.3) [font=\footnotesize]{$p$};
          \node at (2.15,1.6) [elem] {}; \node at (2.15,1.3) [font=\footnotesize]{$q$};

          \draw [dashed,rounded corners=6pt] (0,0) rectangle (1,1);
          \draw [dashed,rounded corners=6pt] (2,0) rectangle (3,1);
          \draw [dashed,rounded corners=6pt] (0,2) rectangle (1,3);
          \draw [dashed,rounded corners=6pt] (2,2) rectangle (3,3);
          
          \node at (0.5,3.5) {$S \cap X$}; \node at (2.5,3.5) {$T \cap X$};
          \node at (0.5,-0.5) {$S \cap Y$}; \node at (2.5,-0.5) {$T \cap Y$};

          \node [dot] at (0.5,2.5) {}; \node at (0.5,2.2) [font=\footnotesize]{$x$};
          \node [dot] at (0.5,0.5) {}; \node at (0.5,0.2) [font=\footnotesize]{$s$};
          \node [dot] at (2.5,0.5) {}; \node at (2.5,0.2) [font=\footnotesize]{$y$};
          \node [dot] at (2.5,2.5) {}; \node at (2.5,2.2) [font=\footnotesize]{$t$};

          \node at (1.5,-1) {(c)};          
        \end{scope}

      \end{tikzpicture}
    \end{center}
    \vspace{-0.35in}
    \caption{Part (a) illustrates the vertex tri-partitions $(S,M,T)$
      and $(X,N,Y)$. \newline In parts (b) and (c), we consider possible
      locations of the terminals $s,t,x,y$.}
  \end{figure}

  We assume w.l.o.g. that $x \in S$. If we also have $y \in S$, then
  $x \in S \cap X$ and $y \in S \cap Y$; therefore, one of $x,y$ is
  diagonally opposite from $t$, suppose this is $x$. Fig. 1 (b)
  illustrates this case. Observe that $A \cup I \cup B$ separates $x$
  from $y$; since $x$ and $y$ are $k_2$-connected and $|N = A \cup I
  \cup C| = k_2$, it follows that $|B| \ge |C|$. Similarly, $C \cup I
  \cup D$ separates $t$ from $s$, and since $C$ contains $q$, Fact 1
  implies that $|C \cup I \cup D| \ge k_1 > |B \cup I \cup D = M| =
  k_1 - 1$. Therefore, $|C| > |B|$, and we have a contradiction.

  Hence, it must be that $y \notin S$; so $y \in T \cap Y$. The
  argument above shows that $x$ and $t$ cannot be diagonally opposite,
  so $t$ must be in $T \cap X$. Similarly, $s$ and $y$ cannot be
  diagonally opposite, so $s \in S \cap Y$. Fig. 1 (c) shows the
  required positions of the vertices. Now, $N$ separates $s$ from $t$
  and contains $p,q$; therefore, from fact 1, $|N| \ge k_1 > |M|$. But
  $M$ separates $x$ from $y$, and fact 2 implies that $x,y$ are
  $k_2$-connected in $G_1$; therefore, $|M| \ge k_2 = |N|$, and we
  have a contradiction.
\end{proof}

\section{Packing Element-Disjoint Steiner Trees and Forests}
\label{sec:packingForests}

Consider a graph $G(V,E)$, with its vertex set $V$ partitioned into
$T_1, T_2, \ldots T_m, W$. We refer to each $T_i$ as a group of
\emph{terminals}, and $W$ as the set of Steiner or white vertices; we
use $T = \bigcup_i T_i$ to denote the set of all terminals. A Steiner
Forest for this graph is a forest that is a subgraph of $G$, such that
each $T_i$ is entirely contained in a single tree of this
forest. (Note that $T_i$ and $T_j$ can be in the same tree.)  For
any group $T_i$ of terminals, we define $\elconn(T_i)$, the
element-connectivity of $T_i$, as the largest $k$ such that for every
$u,v \in T_i$, the element-connectivity of $u$ and $v$ in the graph
$G$ is at least $k$.

We say two Steiner Forests for $G$ are element-disjoint if they share
no edges or Steiner vertices. (Every Steiner Forest must contain all
the terminals.) The Steiner Forest packing problem is to find as many
element-disjoint Steiner Forests for $G$ as possible. By inserting a
Steiner vertex between any pair of adjacent terminals, we can assume
that there are no edges between terminals, and then the problem of
finding element-disjoint Steiner forests is simply that of finding
Steiner forests that do not share any Steiner vertices. A special case
is when $m=1$ in which case we seek a maximum number of element-disjoint
Steiner trees.

\begin{prop} If $k = \min_i \elconn_G(T_i)$, there are at most $k$
  element-disjoint Steiner Forests in $G$.
\end{prop}

Cheriyan and Salavatipour \cite{cs} proved that if there is a single
group $T$ of terminals, with $\elconn(T) = k$, then there always exist
$\Omega(k / \log |T|)$ Steiner trees. Their algorithm proceeds by
using Theorem~\ref{thm:ho}, the global element-connectivity reduction
of \cite{hind}, to delete and contract edges between Steiner vertices,
while preserving $\elconn(T) = k$. Then, once we obtain a bipartite
graph $G'$ with terminals on one side and Steiner vertices on the
other side, randomly color the Steiner vertices using $k/6 \log |T|$
colors; they show that w.h.p., each color class connects the terminal
set $T$, giving $k/ 6 \log |T|$ trees. The bipartite case can be cast
as a special case of packing bases of a polymatroid and a variant of
the random coloring idea is applicable in this more general setting
\cite{CalinescuCV07}; a derandomization is also provided in
\cite{CalinescuCV07}, thus yielding a deterministic polynomial time
algorithm to find $\Omega(k/\log |T|)$ element-disjoint Steiner trees.

In this section, we give algorithms for packing element-disjoint
Steiner Forests, where we are given $m$ groups of terminals $T_1, T_2,
\ldots T_m$. The approach of \cite{cs} encounters two difficulties.
First, we cannot reduce to a bipartite instance, using only the
global-connectivity version of the Reduction Lemma. In fact, our
strengthening of the Reduction Lemma to preserve local connectivity
was motivated by this; using it allows us once again assume that we
have a bipartite graph $G'(T \cup W, E)$.  Second, we cannot apply the
random coloring algorithm on the bipartite graph $G'$ directly; we
give an example in Appendix~\ref{app:packing} to show that this
approach does not work. One reason for this is that, unlike the
Steiner tree case, it is no longer a problem of packing bases of a
submodular function.  To overcome this second difficulty we use a
decomposition technique followed by the random coloring algorithm to
prove that there always exist $\Omega(k/(\log |T| \log m))$
element-disjoint forests.  We believe that the bound can be improved
to $\Omega(k/\log |T|)$.

We also consider the packing problem in restricted classes of graphs,
in particular planar graphs. We obtain a much stronger bound, showing
the existence of $\ceil{k/5} - 1$ Steiner forests. The (simple)
technique extends to graphs of fixed genus to prove the existence of
$\Omega(k)$ Steiner forests where the constant depends mildly on the
genus. We believe that there exist $\Omega(k)$ Steiner forests in any
$H$-minor-free graph where $H$ is fixed; it is shown in \cite{ACJ08}
that there exist $\Omega(k)$ Steiner \emph{trees} in $H$-minor-free
graphs. Our technique for planar graphs does not extend directly, but
generalizing this technique allows us to make partial progress; by
using our general graph result and some related ideas, in
Section~\ref{subsec:treewidth}, we prove that in graphs of any fixed
treewidth, there exist $\Omega(k)$ element-disjoint Steiner Trees if
the terminal set is $k$-element-connected.

\subsection{An $O(\log |T| \log m)$-approximation for Packing
in General Graphs}\label{subsec:generalPacking}

In order to pack element-disjoint Steiner forests we borrow the basic
idea from \cite{ChekuriS} in the {\em edge-connectivity} setting for
Eulerian graphs; this idea was later used by Lau \cite{Lau2} in the
much more difficult non-Eulerian case. The idea at a high level is as
follows: If all the terminals are $k$-connected then we can treat the
terminals as forming one group and reduce the problem to that of
packing Steiner trees. Otherwise, we can find a cut $(S, V\setminus
S)$ that separates some groups from others. If the cut is chosen
appropriately we may be able to treat one side, say $S$, as containing
a single group of terminals and pack Steiner {\em trees} in them {\em
  without} using the edges crossing the cut. Then we can shrink $S$
and find Steiner forests in the reduced graph; unshrinking of $S$ is
possible since we have many trees on $S$. In \cite{ChekuriS,Lau2} this
scheme works to give $\Omega(k)$ edge-disjoint Steiner forests.
However, the approach relies strongly on properties of
edge-connectivity as well as the properties of the packing algorithm
for Steiner trees. These do not generalize easily for
element-connectivity. Nevertheless, we show that the basic idea can be
applied in a slightly weaker way (resulting in the loss of an $O(\log
m)$ factor over the Steiner tree packing factor). We remark that the
reduction to a bipartite instance using the Reduction Lemma plays a
critical role. A key definition is the notion of a good separator
given below.

\begin{definition}
  Given an graph $G(V,E)$ with terminal sets $T_1, T_2, \ldots T_m$,
  such that for all $i$, $\elconn(T_i) \ge k$, we say that a set $S$
  of white vertices is a \emph{good separator} if (i) $|S| \le k/2$
  and (ii) there is a component of $G - S$ in which all terminals are
  $k/2 \log m$-element-connected.
\end{definition}

Note that the empty set is a good separator if all terminals are $k/2
\log m$-element-connected. 

\begin{lemma}\label{lem:separator}
  For any instance of the Steiner Forest Packing problem, there is a
  polynomial-time algorithm that finds a good separator.
\end{lemma}
\begin{proof}
  Let $G(V,E)$ be an instance of the Steiner Forest packing problem,
  with terminal sets $T_1, T_2, \ldots T_m$ such that each $T_i$ is
  $k$-element-connected.  If $T$ is $\frac{k}{2 \log m}$-element
  connected, the empty set $S$ is a good separator.

  Otherwise, there is some set of white vertices of size less than
  $\frac{k}{2 \log m}$ that separates some of the terminals from
  others. Let $S_1$ be a minimal such set, and consider the two or
  more components of $G - S_1$. Note that each $T_i$ is entirely
  contained in a single component, since $T_i$ is at least
  $k$-element-connected, and $|S_1| < k$. Among the components of
  $G-S_1$ that contain terminals, consider a component $G_1$ with the
  fewest sets of terminals; $G_1$ must have at most $m/2$ sets from
  $T_1, \ldots T_m$. If the set of all terminals in $G_1$ is
  $\frac{k}{2 \log m}$ connected, we stop, otherwise, find in $G_1$ a
  set of white vertices $S_2$ with size less than $\frac{k}{2 \log m}$
  that separates terminals of $G_1$.  Again, find a component $G_2$ of
  $G_1 - S_2$ with fewest sets of terminals, and repeat this procedure
  until we obtain some subgraph $G_\ell$ in which all the terminals
  are $\frac{k}{2 \log m}$-connected.  We can always find such a
  subgraph, since the number of sets of terminals is decreasing by a
  factor of $2$ or more at each stage, so we find at most $\log m$
  separating sets $S_j$.  Now, we observe that the set $S = \bigcup_{j
    = 1}^{\ell} S_j$ is a good separator. It separates the terminals
  in $G_\ell$ from the rest of $T$, and its size is at most $\log m
  \times \frac{k}{2 \log m} = k/2$; it follows that each set of
  terminals $T_i$ is entirely within $G_\ell$, or entirely outside
  it. By construction, all terminals in $G_\ell$ are $\frac{k}{2 \log
    m}$ connected.
\end{proof}

We can now prove our main result, that we can always find a packing of
$\Omega(\frac{k}{\log |T| \log m})$ Steiner forests.

\begin{theorem}
  \label{thm:packGeneral}
 Given a graph $G(V,E)$, with terminal sets $T_1, T_2,
  \ldots T_m$, such that for all $i$, $\elconn(T_i) \ge k$, there is a
  polynomial-time algorithm to pack $\Omega(k / \log |T| \log m)$
  element-disjoint Steiner Forests in $G$.
\end{theorem}
\begin{proof}
  The proof is by induction on $m$. The base case of $m = 1$, follows
  from \cite{cs,CalinescuCV07}; $G$ contains at least $\frac{k}{6 \log |T|}$
  element-disjoint Steiner \emph{Trees}, and we are done.

  We may assume $G$ is bipartite by using the Reduction Lemma. Find a
  good separator $S$, and a component $G_\ell$ of $G-S$ in which all
  terminals are $\frac{k}{2 \log m}$-connected.  Now, since the
  terminals in $G_\ell$ are $\frac{k}{2 \log m}$-connected, use the
  algorithm of \cite{cs} to find $\frac{k}{12 \log m \log |T|}$
  element-disjoint Steiner trees containing all the terminals in
  $G_\ell$; none of these trees uses vertices of $S$.  Number these
  trees from 1 to $\frac{k}{12 \log m \log |T|}$; let $\script{T}_j$
  denote the $j$th tree.

  The set $S$ separates $G_\ell$ from the terminals in $G -
  G_\ell$. If $S$ is not a minimal such set, discard vertices until it
  is. If we delete $G_\ell$ from $G$, and add a clique between the
  white vertices in $S$ to form a new graph $G'$, it is clear that the
  element-connectivity between any pair of terminals in $G'$ is at
  least the element-connectivity they had in $G$.  The graph $G'$ has
  $m' \le m-1$ groups of terminals; by induction, we can find
  $\frac{k}{12 \log |T| \log m} < \frac{k}{12 \log |T| \log m'}$
  element-disjoint Steiner forests for the terminals in $G'$. As
  before, number the forests from 1 to $\frac{k}{12 \log m \log |T|}$;
  we use $\script{F}_j$ to refer to the $j$th forest. These Steiner
  Forests may use the newly added edges between the vertices of $S$;
  these edges do not exist in $G$.  However, we claim that the Steiner
  Forest $\script{F}_j$ of $G'$, together with the Steiner tree
  $\script{T}_j$ in $G_\ell$ gives a Steiner Forest of $G$. The only
  way this might not be true is if $\script{F}_j$ uses some edge added
  between vertices $u,v \in S$. However, every vertex in $S$ is
  adjacent to a terminal in $G_\ell$, and all the terminals of
  $G_\ell$ are in every one of the Steiner trees we
  generated. Therefore, there is a path from $u$ to $v$ in
  $\script{T}_j$. Hence, deleting the edge between $u$ and $v$ from
  $\script{F}_j$ still leaves each component of $\script{F}_j \cup
  \script{T}_j$ connected.

  Therefore, for each $1 \le j \le \frac{k}{12 \log m \log |T|}$, the
  vertices in $\script{F}_j \cup \script{T}_j$ induce a Steiner Forest
  for $G$.
\end{proof}

\subsection{Packing Steiner Trees and Forests in Planar Graphs}
\label{subsec:planarPacking}

We now prove much improved results for restricted classes of graphs,
in particular planar graphs. If $G$ is planar, we show the existence
of $\ceil{k/5} - 1$ element-disjoint Steiner Forests.\footnote{Note
  that in the special case of packing Steiner Trees, the paper of
  Aazami \etal \cite{ACJ08} shows that there are $\floor{k/2} - 1$
  element-disjoint Steiner Trees.} The intuition and algorithm are
easier to describe for the Steiner tree packing problem and we do this
first. We achieve the improved bound by observing that planarity
restricts the use of many white vertices as ``branch points'' (that
is, vertices of degree $\ge 3$) in forests. Intuitively, even in the
case of packing trees, if there are terminals $t_1, t_2, t_3, \ldots $
that must be in every tree, and white vertices $w_1, w_2, w_3 \ldots$
that all have degree 3, it is difficult to avoid a $K_{3,3}$
minor. Note, however, that degree $2$ white vertices behave like edges
and do not form an obstruction.  We capture this intuition more
precisely by showing that there must be a pair of terminals $t_1,t_2$
that are connected by $\Omega(k)$ degree-2 white vertices; we can
contract these ``parallel edges'', and recurse.

We describe below an algorithm for packing Steiner Trees.  Through the
rest of the section, we assume $k > 10$; otherwise, $\ceil{k/5} - 1
\le 1$, and we can always find 1 Steiner Tree in a connected graph.

Given an instance of the Steiner Tree packing problem in planar
graphs, we construct a \emph{reduced instance} as follows: Use the
Reduction Lemma to delete and contract edges between white vertices to
obtain a planar graph with vertex set $T \cup W$, such that $W$ is a
stable set. Now, for each vertex $w \in W$ of degree 2, connect the
two terminals that are its endpoints directly with an edge, and delete
$w$. (All edges have unit capacity.)  We now have a planar
\emph{multigraph}, though the only parallel edges are between
terminals, as these were the only edges added while deleting degree-2
vertices in $W$. Note that this reduction preserves the
element-connectivity of each pair of terminals; further, any set of
element-disjoint trees in this reduced instance corresponds to a set
of element-disjoint trees in the original instance. We need the
following technical result:

\begin{theorem}[Borodin, \cite {Borodin}]
  \label{thm:Borodin}
  If $G$ is a planar graph with minimum degree $3$, it has an edge of
  weight at most $13$, where the weight of an edge is the sum of the
  degrees of its endpoints.
\end{theorem}

\begin{lemma} \label{lem:parallelEdges} 
  In a reduced instance of the Planar Steiner Tree Packing problem, if
  $T$ is $k$-element-connected, there are two terminals $t_1, t_2$
  with at least $\ceil{k/5} - 1$ parallel edges between them.
\end{lemma}

\begin{proof}
  We prove this lemma in Appendix~\ref{app:planarPacking}; here, we
  give a proof showing the weaker result that there exist terminals
  $t_1, t_2$ with $\ceil{k/10}$ edges between them. Let $G$ be the
  planar multigraph of the reduced instance. Since $T$ is
  $k$-element-connected in $G$, every terminal has degree at least $k$
  in $G$. Construct a planar graph $G'$ from $G$ by keeping only a
  single copy of each edge. We argue below that some terminal $t_1 \in
  T$ has degree at most $10$ in $G'$; it follows that $G$ must contain
  at least $\ceil{k/10}$ copies of some edge incident to $t_1$, as
  $t_1$ has degree at least $k$ in $G$. These edges must be incident
  to another terminal $t_2$, completing the proof.

  To see that some terminal $t_1$ has degree at most $10$ in $G'$, we
  first assume that no terminal has degree $\le 2$, or we are already
  done. Now, as every vertex of $W$ in a reduced instance has degree
  at least $3$, we may use Theorem~\ref{thm:Borodin}; this implies
  that $G'$ has an edge $e$, such that the sum of the degrees of the
  endpoints of $e$ is at most 13. The edge $e$ must be incident to a
  terminal $t_1$, as the white vertices are a stable set. The other
  endpoint of $e$ has degree at least $3$, so the degree of $t_1$ is
  at most $10$.
\end{proof}

It is now easy to prove by induction that we can pack $\ceil{k/5} -
1$ disjoint trees.

\begin{theorem}\label{thm:planarTreePacking}
  Given an instance of the Steiner Tree packing problem on a planar
  graph $G$ with terminal set $T$, if $\elconn(T) \ge k$, there is a
  polynomial-time algorithm to find at least $\ceil{k/5} - 1$
  element-disjoint Steiner trees in $G$. Moreover, in each tree, 
  the white (non-terminal) vertices all have degree $2$.
\end{theorem}

\begin{proof}
  We prove this theorem by induction on $|T|$; if $|T| = 2$, there are
  $k$ disjoint \emph{paths} in $G$ from one terminal to the other, so
  we are done (including the guarantee of degree $2$ for white
  vertices).

  Otherwise, apply the Reduction Lemma to construct a reduced instance
  $G'$, preserving the element-connectivity of $T$. Now, from
  Lemma~\ref{lem:parallelEdges}, there exist a pair of terminals $t_1,
  t_2$ that have $\ceil{k/5} - 1$ parallel edges between them (Note
  that the parallel edges between $t_1$ and $t_2$ may have
  non-terminals on them in the original graph but they have degree
  $2$.). Contract $t_1, t_2$ into a single terminal $t$, and consider
  the new instance of the Steiner Tree packing problem with terminal
  set $T' = T \cup \{t\} - \{t_1, t_2\}$. It is easy to see that the
  element-connectivity of the terminal set is still at least $k$; by
  induction, we can find $\ceil{k/5} - 1$ Steiner trees containing all
  the terminals of $T'$, with the property that all non-terminals have
  degree $2$. Taking these trees together with $\ceil{k/5} - 1$ edges
  between $t_1$ and $t_2$ gives $\ceil{k/5} - 1$ trees in $G'$ that
  span the original terminal set $T$.
\end{proof}

\smallskip
\noindent {\bf Packing Steiner Forests in Planar Graphs:} The
algorithm described above for packing Steiner trees encounters a
technical difficulty when we try to extend it to Steiner forests.
Lemma~\ref{lem:parallelEdges} can be used at the start to merge some
two terminals. However, as the algorithm proceeds it may get stuck in
the following situation: it merges all terminals from some group $T_i$
into a single terminal. Now this terminal does not require any more
connectivity to other terminals although other groups are not yet
merged together. In this case we term this terminal as dead. In the
presence of dead terminals Lemma~\ref{lem:parallelEdges} no longer
applies; we illustrate this with a concrete example in
Appendix~\ref{subsec:forestPacking}.  We overcome this difficulty by
showing that a dead terminal may be replaced by a grid of white
vertices --- the grid is necessary to ensure that the resulting graph
is still planar.  We can then apply the Reduction Lemma to remove
edges between the newly added white vertices and proceed with the
merging process. See Appendix~\ref{subsec:forestPacking} for details.

\smallskip
\noindent {\bf Extensions:} Our result for planar graphs can be
generalized to graphs of fixed genus; Ivanco \cite{Ivanco} generalized
Theorem~\ref{thm:Borodin} to show that a graph $G$ of genus $g$ has an
edge of weight at most $2g + 13$ if $0 \le g \le 3$ and an edge of
weight at most $4g + 7$ otherwise.  This allows us to prove that there
exist $\ceil{k/c}$ forests where $c \le 4g+8$; we have not attempted
to optimize this constant $c$.  Aazami \etal \cite{ACJ08} also give
algorithms for packing Steiner Trees in these graph classes, and
graphs excluding a fixed minor. We thus make the following natural
conjecture:

\begin{conjecture}
  \label{conj:minorfree}
  Let $G=(V,E)$ be a $H$-minor-free graph, with terminal sets $T_1,
  T_2, \ldots T_m$, such that for all $i$, $\elconn(T_i) \ge k$.
  There exist $\Omega(k/c)$ element-disjoint Steiner forests in $G$,
  where $c$ depends only on the size of $H$.
\end{conjecture}

We note that Lemma~\ref{lem:parallelEdges} fails to hold for
$H$-minor-free graphs, and in fact fails even for bounded treewidth
graphs. Thus, our approach cannot be directly generalized. However,
instead of attempting to contract together just two terminals
connected by many parallel edges, we may be able contract together a
constant number of terminals that are ``internally'' highly
connected. Using Theorem~\ref{thm:packGeneral} and other ideas, we
prove in the next section that this approach suffices to pack many
trees in graphs with small treewidth. We believe that these ideas
together with the structural characterization of $H$-minor-free graphs
by Robertson and Seymour \cite{RobertsonSeymour} should lead to a
positive resolution of Conjecture~\ref{conj:minorfree}.

\subsection{Packing Trees in Graphs of Bounded Treewidth}
\label{subsec:treewidth}

Let $G(V,E)$ be a graph of treewidth $\le r-1$, with terminal set $T
\subseteq V$ such that $\elconn(T) \ge k$. In this section, we give an
algorithm to find, for any fixed $r$, $\Omega(k)$ element-disjoint
Steiner Trees in $G$. Our approach is similar to that for packing
Steiner Trees in planar graphs, where we argued in
Lemma~\ref{lem:parallelEdges} that there exist two terminals $t_1,
t_2$ with $\Omega(k)$ parallel edges between them, so we could
contract them together and recurse on a smaller instance. In graphs of
bounded treewidth, this is no longer the case; see the end of
Appendix~\ref{app:packing} for an example in which no pair of
terminals is connected by many parallel edges.  However, we argue that
there exists a small set of terminals $T' \subset T$ that is highly
``internally connected'', so we can find $\Omega(k)$ disjoint trees
connecting all terminals in $T'$, without affecting the connectivity
of terminals in $T - T'$. We can then contract together $T'$ and the
white vertices used in these trees to form a single new terminal $t$,
and again recurse on a smaller instance. The following lemma captures
this intuition:

\begin{lemma}\label{lem:treewidthSeparator}
  If $G(V, E)$ is a bipartite graph of treewidth at most $r-1$, with
  terminal set $T \subset V$ such that $T \ge 2^r$, $\elconn(T) \ge
  k$, there exists a set $S \subseteq V - T$ such that there is a
  component $G'$ of $G - S$ containing $k/12 r^2 \log (3r)$
  element-disjoint Steiner trees for the (at least 2) terminals in
  $G'$. Moreover, these trees in $G'$ can be found in polynomial time.
\end{lemma}

Given this lemma, we prove below that for any fixed $r$, we can pack
$\Omega(k)$ element-disjoint trees in graphs of treewidth at most
$r-1$. The proof combines ideas of Theorem~\ref{thm:planarTreePacking}
and Theorem~\ref{thm:packGeneral}.

\begin{theorem} \label{thm:treewidth}
  Let $G=(V,E)$ be a graph of treewidth at most $r-1$.  For any terminal
  set $T \subseteq V$ with $\elconn_G(T) \ge k$, there exist
  $\Omega(k/12 r^2 \log (3r))$ element-disjoint Steiner trees on $T$.
\end{theorem}
\begin{proof}
  As for Theorem~\ref{thm:planarTreePacking}, we prove this theorem by
  induction. Let $G$ be a graph of treewidth at most $r-1$, with
  terminal set $T$. If $|T| \le 2^r$, we have $k/6 \log |T| \ge k/6r$
  element-disjoint trees from the tree-packing algorithm of Cheriyan
  and Salavatipour \cite{cs} in \emph{arbitrary} graphs.

  Otherwise, we use the Reduction Lemma to ensure that $G$ is
  bipartite. Let $S$ be a set of white vertices guaranteed to exist
  from Lemma~\ref{lem:treewidthSeparator}. If $S$ is not a minimal
  such set, discard vertices until it is. Now, find $k/12 r^2 \log
  (3r)$ element-disjoint trees containing all terminals in some
  component $G'$ of $G - S$; note that each vertex of $S$ is incident
  to some terminal in $G'$, and hence to every tree. (This follows
  from the minimality of $S$ and the fact that $G$ is bipartite.)
  Modify $G$ by contracting all of $G'$ to a single terminal $t$, and
  make it incident to every vertex of $S$. It is easy to see that all
  terminals in the new graph are $k$-element-connected; therefore, we
  now have an instance of the Steiner Tree packing problem on a graph
  with fewer terminals. The new graph has treewidth at most $r-1$, so
  by induction, we have $k/12 r^2 \log (3r)$ element-disjoint trees
  for the terminals in this new graph; taking these trees together
  with the $k / 12r^2 \log (3r)$ trees of $G'$ gives $k/12 r^2 \log
  (3r)$ trees of the original graph $G$.
\end{proof}

We devote the rest of this section to proving the crucial
Lemma~\ref{lem:treewidthSeparator}.  Subsequently, we may assume,
w.l.o.g. (after using the Reduction Lemma) that the graph $G$ is
bipartite; we may further assume that $k \ge 12r^2 \log (3r)$ and $|T|
\ge 2^r$. First, observe that $G$ has a small cutset that separates a
few terminals from the rest.

\begin{prop}
  $G$ has a cutset $C$ of size at most $r$ such that some component of
  $G - C$ contains between $r$ and $2r$ terminals.
\end{prop}
\begin{proofsketch}
  Fix a tree-decomposition $\script{T}$ of $G$; every non-leaf node of
  $\script{T}$ corresponds to a cutset, and each node of $\script{T}$
  contains at most $r$ vertices of $G$. Start at a leaf of
  $\script{T}$, and walk upwards until reaching a node $v$ such that
  the subtree of $\script{T}$ rooted at some child of $v$ contains
  between $r$ and $2r$ terminals. (This is always possible since
  walking up one step only gives at most $r$ more terminals.)
\end{proofsketch}

We find the set $S$ and component of $G - S$ in which we contract
together a small number of terminals by focusing on the cutset $C$ and
component of $G-C$ that are guaranteed to exist from the previous
proposition. We introduce some notation before proceeding with the proof:

\begin{enumerate}
\item Let $C$ be a cutset of size at most $r$, and let $V'$ be the
  vertices of a component of $G - C$ containing between $r$ and $2r$
  terminals.

\item Since terminals in $V'$ are $k$-connected to the terminals in
  the rest of the graph, and $|C| \le r \ll k$, $C$ contains at least
  one black vertex. Let $C'$ be the set of black vertices in $C$.

\item Let $G' = G[V' \cup C']$ be the graph induced by $V'$ and $C'$.
\end{enumerate}

We omit a proof of the following straightforward proposition; the
second part of the statement follows from the fact that each terminal in
$V'$ is $k$-connected to terminals outside $G'$, and these paths to
terminals outside $G'$ must go through the cutset $C'$ of size at most
$r$.

\begin{prop}\label{prop:V'Connected}
  The graph $G'$ contains between $r$ and $3r$ terminals (as $C'$ may
  contain up to $r$ terminals), and each terminal in $V'$ is at least
  $k/r$-connected to some terminal in $C'$. 
\end{prop}

Let $T'$ be the set of terminals in $G'$. If $\elconn_{G'}(T') \ge
k/2r^2$, we can easily find a set of white vertices satisfying
Lemma~\ref{lem:treewidthSeparator}: Let $S$ be the set of vertices of
$G$ that are adjacent (in $G$) to vertices of $G'$.  It is obvious
that $S$ separates $G'$ from the rest of $G$, and all terminals in
$T'$ are highly connected; from the tree packing result of \cite{cs},
we can find the desired disjoint trees in $G'$. Finally, note that all
vertices of $S$ are white, as the only neighbors of $G'$ are either
white vertices of the cutset $C$ or the neighbors of the black
vertices in $C$, all of which are white as $G$ is bipartite.

However, it may not be the case that all terminals of $T'$ are highly
connected in $G'$. In this event, we use the following simple
algorithm (very similar to that in the proof of
Lemma~\ref{lem:separator}) to find a highly-connected subset of $T'$:
Begin by finding a set $S_1$ of at most $k/2r^2$ white vertices in
$G'$ that separates terminals of $T'$. Among the components of $G' -
S_1$, pick a component $G_1$ with at least one terminal of $V'$. If
all terminals of $G_1$ are $k/2r^2$ connected, stop; otherwise, find
in $G_1$ a set $S_2$ of at most $k/2r^2$ white vertices that separates
terminals of $G_1$, pick a component $G_2$ of $G_1 - S_2$ that
contains at least one terminal of $V'$, and proceed in this manner
until finding a component $G_\ell$ in which all terminals are $k/2r^2$
connected.

\begin{claim}
  We perform at most $r$ iterations of this procedure before we stop,
  having found some subgraph $G_\ell$ in which all the (at least 2)
  terminals are $k/2r^2$ connected.
\end{claim}
\begin{proof}
  At least one terminal of $C'$ must be lost every time we find such a
  set $S_i$; if this is true, the claim follows. To see that this is
  true, observe that when we find a cutset $S_{i+1}$ in $G_i$, there
  is a component that we do \emph{not} pick that contains a terminal
  $t$. If this terminal $t$ is in $C'$, we are done; otherwise, it
  must be in $V'$. But from Proposition~\ref{prop:V'Connected} all
  terminals in $V'$ are $k/r$ connected to some terminal in $C'$, and
  so some terminal of $C'$ must be in the same component as $t$. When
  we stop with the subgraph $G_\ell$, it contains at least one
  terminal $t' \in V'$, and at least one terminal of $C'$ to which
  $t'$ is highly connected; therefore, $G_\ell$ contains at least 2
  terminals.
\end{proof}

All terminals in the subgraph $G_\ell$ are $k/2r^2$-connected, and
there are at most $3r$ of them, so we can find $k/12 r^2 \log (3r)$
disjoint trees \emph{in $G_\ell$} that connect them, using the
tree-packing result of \cite{cs}. Let $S$ be the set of vertices of
$G$ that are adjacent (in $G$) to vertices of $G_\ell$; obviously, $S$
separates $G_\ell$ from the rest of $G$, and to satisfy
Lemma~\ref{lem:treewidthSeparator}, it merely remains to verify that
$S$ only contains white vertices. Every terminal in $G' - G_\ell$ was
separated from $G_\ell$ by white vertices in some $S_i$, and terminals
in $G - G'$ can only be incident to white vertices of the cutset $C$,
which are not in $G'$, let alone $G_\ell$. This completes the proof of
Lemma~\ref{lem:treewidthSeparator}.

\section{Single-Sink Vertex-Connectivity}\label{sec:kconn}
Recall that in the \sskconn~problem, one is given an undirected graph
$G=(V,E)$ with edge costs, a specified sink/root vertex $r$, and a
subset of terminals $T \subseteq V$, with $|T| = h$.  The goal is to
find a minimum cost subgraph $H$ that contains $k$ vertex-disjoint
paths from each terminal $t \in T$ to the root. In this section we
give a very simple proof of the main technical result in
\cite{ChuzhoyK08} using the Reduction Lemma. We lead up to the technical
lemma via a description of the (simple) algorithm for \sskconn.

The basic algorithmic idea comes from \cite{ChakCK08}; this is the
idea of using \emph{augmentation}. Let $T' \subseteq T$ be a subset of
terminals and let $H'$ be a subgraph of $G$ that is feasible for
$T'$. For a terminal $t \in T \setminus T'$, a set of $k$ paths $p_1,
\ldots, p_k$ is said to be an augmentation for $t$ with respect to
$T'$ if (i) $p_i$ is a path from $t$ to some vertex in $T' \cup \{r\}$
(ii) the paths are internally vertex disjoint and (iii) a terminal $t'
\in T'$ is the endpoint of at most one of the $k$ paths. Note that the
root is allowed to be the endpoint of more than one path. The
following proposition is easy to prove via a simple min-cut argument.

\begin{prop}
  \label{prop:aug}
  If $p_1, p_2, \ldots, p_k$ is an augmentation for $t$ with respect
  to $T'$ and $H'$ is a feasible solution for the \sskconn\ instance
  with terminal set $T'$, then $H \cup (\bigcup_i p_i)$ is a feasible
  solution for $T' \cup \{t\}$.
\end{prop}

Given $T'$ and $t$, the augmentation cost of $t$ with respect to $T'$
is the cost of a min-cost set of paths that augment $t$ w.r.t. to
$T'$. We can find the augmentation cost for a terminal $t$ by solving
a simple min-cost flow problem. The key theorem in
\cite{ChuzhoyK08} is the following.

\begin{theorem}[Vertex-Connectivity, \cite{ChuzhoyK08}] 
  \label{thm:kconnAug}
  If $\opt$ denotes the cost of an optimal solution to \sskconn, and
  $AugCost(t)$ the cost of an augmentation for terminal $t$ w.r.t. $T
  - \{t\}$, $\sum_t AugCost(t) \le 8k \cdot \opt$.
\end{theorem}

We now briefly describe the algorithm of \cite{ChekuriK08} for
\sskconn; a variant is used in \cite{ChakCK08, ChuzhoyK08}. 

\vspace{-0.1in}
\begin{algo}
  Permute the terminals randomly; let $t_j$ denote the $j$th terminal in
  the permutation and let $T_j = \{t_1, \ldots, t_j\}$. \\
  Subgraph $H \leftarrow \emptyset$\\
  For $i = 1$ to $|T|$. \+ \\
    Add to $H$ a min-cost augmentation of $t_i$ with respect to
    $T_{i-1}$. \- \\
  Output the subgraph $H$.
\end{algo}

Note that the above is a greedy algorithm except for the initial
randomization. Interestingly, as noted in \cite{ChekuriK08}, the
randomization is key; even for $k=2$ there exist permutations that
yield a solution of cost $\Omega(|T| \cdot \opt)$. Using
Theorem~\ref{thm:kconnAug} it is easy to prove that the above
algorithm is a randomized $O(k \log |T|)$-approximation for \sskconn:
simply observe that the \emph{expected} augmentation cost for the last
terminal in the permutation is at most $8k \opt/|T|$; a
straightforward inductive argument then completes the proof. 

The main ingredient in the proof of Theorem~\ref{thm:kconnAug}, as
shown by \cite{ChuzhoyK08}, is the following weaker statement
involving paths that are \emph{element-disjoint}, as opposed to
vertex-disjoint.

\begin{lemma}[Element-Connectivity, \cite{ChuzhoyK08}]
  \label{lem:elemConnectivity}
  Given an instance of \sskconn, let $ElemCost(t)$ denote the minimum
  cost of a set of $k$ internally vertex-disjoint paths from any terminal
  $t$ to $T \cup \{r\} - t$. Then, $\sum_{t \in T} ElemCost(t) \le 2
  \opt$, where $\opt$ is the cost of an optimal solution to this
  instance.
\end{lemma}

It is shown in \cite{ChuzhoyK08} that one can prove
Theorem~\ref{thm:kconnAug} by repeatedly invoking
Lemma~\ref{lem:elemConnectivity} to obtain a large collection of paths
from each $t \in T$ to other terminals, and applying a flow-scaling
argument. The heart of the proof of the crucial
Lemma~\ref{lem:elemConnectivity}, is a structural theorem of
\cite{ChuzhoyK08} on \emph{spiders}: A spider is a tree containing at
most a single vertex of degree greater than 2. If such a vertex
exists, it is referred to as the \emph{head} of the spider, and each
leaf is referred to as a \emph{foot}.  Thus, a spider may be viewed as
a collection of disjoint paths (called \emph{legs}) from its feet to
its head. If the spider has no vertex of degree 3 or more, any vertex
of the spider may be considered its head.  Vertices that are not the
head or feet are called intermediate vertices of the spider. The
Reduction Lemma allows us to give an extremely easy inductive proof of
the Spider Decomposition Theorem below,\footnote{In the decomposition
  theorem of \cite{ChuzhoyK08}, the spiders satisfy a certain
  additional technical condition; the proof of
  Theorem~\ref{thm:kconnAug} in \cite{ChuzhoyK08} relies on this
  condition. We give a modified proof of Theorem~\ref{thm:kconnAug}
  that does not require the condition.}
greatly simplifying the proof of \cite{ChuzhoyK08}.

\begin{theorem}[\cite{ChuzhoyK08}]\label{thm:spiders}
  Let $G(V, E)$ be a graph with a set $B \subseteq V$ of black
  vertices such that every pair of black vertices is $k$-element
  connected. There is a subgraph $H$ of $G$ whose edges can be
  partitioned into spiders such that:
  \vspace{-0.1in}
  \begin{enumerate}
    \item For each spider, its feet are distinct black vertices, and
      all intermediate vertices are white.

    \item Each black vertex is a foot of exactly $k$ spiders, and each
      white vertex appears in at most one spider. 

    \item If a white vertex is the head of a spider, the spider has at
      least two feet.
  \end{enumerate}
\end{theorem}

Before giving the formal short proof we remark that if the graph is
bipartite then the collection of spiders is trivial to see: they are
simply the edges between the black vertices and the stars rooted at
each white vertex! Thus the Reduction Lemma effectively
allows us to reduce the problem to a trivial case. ~\\

\begin{proof}
  We prove this theorem by induction on the number of edges between
  white vertices in $G$. As the base case, we have a graph $G$ with no
  edges between white vertices; therefore, $G$ is bipartite. (Recall
  that there are no edges between black vertices.)  Each pair of black
  vertices is $k$-element connected, and hence every black vertex has
  at least $k$ white neighbors. Let every $b \in B$ mark $k$ of its
  (white) neighbors arbitrarily. Every white vertex $w$ that is marked
  at least twice becomes the head of a spider, the feet of which are
  the black vertices that marked $w$. For each white vertex $w$ marked
  only once, let $b$ be its neighbor that marked it, and $b'$ be
  another neighbor. We let $b-w-b'$ be a spider with foot $b$ and head
  $b'$. It is easy to see that the spiders are disjoint, and that they
  satisfy all the other desired conditions.

  For the inductive step, consider a graph $G$ with an edge $pq$
  between white vertices. If all black vertices are $k$-element
  connected in $G_1 = G - pq$, then we can apply induction, and find
  the desired subgraph of $G_1$ and hence of $G$. Otherwise, by
  Theorem~\ref{lem:reduction}, we can find the desired set of spiders
  in $G_2 = G / pq$. If the new vertex $v = pq$ is not in any spider,
  this set of spiders exists in $G$, and we are done. Otherwise, let
  $S$ be the spider containing $v$. If $v$ is not the head of $S$, let
  $x,y$ be its neighbors in $S$. Either $x$ and $y$ are both adjacent
  to $p$, or both adjacent to $q$, or (w.l.o.g.) $x$ is adjacent to
  $p$ and $y$ to $q$. Therefore, we can replace the path $x-v-y$ in
  $S$ with one of $x-p-y$, $x-q-y$, or $x-p-q-y$.  If $v$ is the head
  of $S$, we know that it has at least 2 feet. If at least 2 legs of
  $S$ are incident to each of $p$ and $q$, we can create two new
  spiders $S_p$ and $S_q$, with heads $p$ and $q$ respectively; $S_p$
  contains the legs of $S$ incident to $p$, and $S_q$ the legs
  incident to $q$. If all the legs of $S$ are incident to $p$, we let
  $p$ be the head of the spider in $G$; the case in which all legs are
  incident to $q$ is symmetric. If neither of these cases holds, it
  follows that (w.l.o.g.)  exactly one leg $\ell$ of $S$ is incident
  to $p$, with the remaining legs being incident to $q$. We let $q$ be
  the head of the new spider, and add $p$ to the leg $\ell$.
\end{proof}

The authors of \cite{ChuzhoyK08} showed that, once we have the Spider
Decomposition Theorem, it is very easy to prove
Lemma~\ref{lem:elemConnectivity}.

\begin{proofof}{Lemma~\ref{lem:elemConnectivity}}(\cite{ChuzhoyK08})
  In an optimal solution $H$ to an instance of \sskconn, every
  terminal is $k$-vertex-connected to the root. Let the terminals be
  black vertices, and non-terminals be white; it follows that all the
  terminals are $k$-element connected to the root in $H$, and hence to
  each other. Therefore, we can find a subgraph of $H$ of total cost
  at most $\opt$ which can be partitioned into spiders as in
  Theorem~\ref{thm:spiders}. For each spider $S$ and every terminal
  $t$ that is a foot of $S$, we find a path entirely contained within
  $S$ from $t$ to another terminal. Each edge of $S$ is in at most two
  such paths; since the spiders are disjoint and each terminal is a
  foot of $k$ spiders, we obtain the desired result.

  If the head of $S$ is a terminal, the path for each foot is simply
  the leg of $S$ from that foot to the head. Each edge of $S$ is in a
  single path. If the head of $S$ is a white vertex, it has at least
  two feet. Fix an arbitrary ordering of the feet of $S$; the path for
  foot $i$ follows leg $i$ from the foot to the head, and then leg
  $i+1$ from the head to foot $i+1$. (The path for the last foot
  follows the last leg, and then leg 1 from the head to the foot.)
  It is easy to see that each edge of $S$ is in exactly two paths;
  this completes the proof.
\end{proofof}

Finally, we give a proof of Theorem~\ref{thm:kconnAug} that relies
only on the statement of Lemma~\ref{lem:elemConnectivity}. Our
proof is a technical modification of the one in \cite{ChuzhoyK08} and
as previously remarked, does not need rely on the additional condition
on the spiders that \cite{ChuzhoyK08} guarantees. Our proof
also gives a slightly stronger bound on $\sum_t AugCost(t)$ ($8k \cdot
\opt$ instead of $(18 k + 3)\cdot \opt$).

\begin{proofof}{Theorem~\ref{thm:kconnAug}}
  We give an algorithm to find an augmentation for each terminal that
  proceeds in $4k^2$ iterations: In each iteration, for every terminal
  $t$, it finds a set of $k$ internally vertex-disjoint paths from $t$
  to other terminals or the root. Let $\script{P}_i(t)$ denote the set
  of paths found for terminal $t$ in iteration $i$. These paths have
  the following properties:
  \begin{enumerate}
    \item For each terminal $t$, every other terminal is an end-point
      of fewer than $4k^2 + 2k$ paths in $\bigcup_i \script{P}_i(t)$.
      
    \item In each iteration $i$, $\sum_t Cost(\script{P}_i(t)) \le 4k
      \opt$.
  \end{enumerate}

  Given these two properties, we can prove the theorem as follows:
  Separately for each terminal $t$, send 1 unit of flow along each of
  the paths in $\bigcup_i \script{P}_i(t)$; we thus have a flow of
  $4k^2 \cdot k$ units from $t$ to other terminals. Scale this flow
  down by $4k^2 \cdot (k+\frac{1}{2})/k$, to obtain a flow of
  $\frac{k^2}{k + 1/2} > k-1/2$ from $t$ to other terminals.  After
  the scaling step, the net flow through any vertex (terminal or
  non-terminal) is at most 1, since the maximum flow through a vertex
  before scaling was $4k^2 + 2k$. Let $FlowCost(t)$ denote the cost of
  this scaled flow for terminal $t$; if we now scale the flow
  \emph{up} by a factor of 2, we obtain a flow of value greater than
  $2k-1$ from $t$ to other terminals, in which the flow through any
  vertex besides $t$ is at most 2. Therefore, by the integrality of
  min-cost flow, we can find an integral flow of $2k-1$ units from $t$
  to other terminals, of total cost at most $2 FlowCost(t)$. Let $E_t$
  be the set of edges used in this integral flow; it follows that
  $cost(E_t) \le 2FlowCost(t)$. It is also easy to see that $E_t$
  contains $k$ disjoint paths from $t$ to $k$ distinct terminals, by
  observing that a hypothetical cutset of size $k-1$ contradicts the
  existence of the flow of value $2k-1$ in which the flow through a
  vertex is at most 2.

  Therefore, we have found $k$ disjoint paths from $t$ to $k$ other
  terminals, of total cost $2 FlowCost(t)$. To bound the cost over all
  terminals, we note that from the second property above, we have
  $\sum_t FlowCost(t) \le 4k^2 \cdot 4k \opt / \left( 4k^2
    \frac{k+1/2}{k} \right)$, which is less than $4k \opt$. It follows
  that the total cost of the set of paths is at most $2 \sum_t
  FlowCost(t) < 8k \opt$.

  \bigskip 
  It remains only to show that we can find a set of paths for each
  terminal in every iteration that satisfies the two desired
  properties.  The proof below uses induction on the number of
  iterations $i$ to prove property 1: After $i$ iterations, for each
  terminal $t$, every other terminal is an end-point of fewer than $i
  + 2k$ paths in $\bigcup_i \script{P}_i(t)$.

  In iteration $i$, for each terminal $t$, let $Blocked(t)$ denote the
  set of terminals in $T - t$ that have been the endpoints of at least
  $(i-1) + k$ paths in $\bigcup_{j=1}^{i-1} \script{P}_j(t)$. (Note
  that the root $r$ is never in any $Blocked(t)$.) Since the total
  number of paths that have been found so far is $(i-1)k$,
  $|Blocked(t)| < k$. Construct a directed graph $D$ on the set of
  terminals, with edges from each terminal $t$ to the terminals in
  $Blocked(t)$. Since the out-degree of each vertex in $D$ is at most
  $k-1$, there is a vertex of in-degree at most $k-1$; therefore, the
  digraph $D$ is $2k-2$ degenerate and so can be colored using $2k-1$
  colors. Let $C_1, C_2, \ldots C_{2k-1}$ denote the color classes in
  a proper coloring of $D$; if $t_1, t_2 \in C_j$, then in iteration
  $i$, $t_1 \notin Blocked(t_2)$ and $t_2 \notin Blocked(t_1)$. For
  each color class $C_j$ in turn, consider the terminals of $C_j$ as
  black, and the non-terminals and terminals of other classes as
  white. There is a graph of cost $\opt$ in which every terminal of
  $C_j$ is $k$-vertex-connected to the root, so $C_j$ is
  $k$-element-connected to the root in this graph even if terminals
  not in $C_j$ are regarded as white vertices. From
  Lemma~\ref{lem:elemConnectivity}, for every $C_j$, we can find a set
  of internally disjoint paths from each $t \in C_j$ to $C_j \cup\{r\}
  - \{t\}$ of total cost at most $2 \opt$. If these paths contain
  other terminals in $T - C_j$ as intermediate vertices, trim them at
  the first terminal they intersect. It follows that $\sum_j \sum_{t
    \in C_j} Cost(\script{P}_i(t)) < 4k \opt$, establishing property 2
  above.

  To conclude, we show that for each terminal $t$, after iteration
  $i$, every other terminal is an end-point of fewer than $i + 2k$
  paths in $\bigcup_{j=1}^i \script{P}_j(t)$. Let $C$ be the color
  class containing $t$; if $t' \in Blocked(t)$, at most one new path
  in $\script{P}_i(t)$ ends in $t'$, as the paths for $t$ are disjoint
  except at terminals in $C$, and $t' \notin C$. By induction, before
  this iteration $t'$ was the endpoint of fewer than $(i-1) + 2k$
  paths for $t$, and so after this iteration, it cannot be the
  endpoint of $i+2k$ paths for $t$. If $t' \notin Blocked(t)$, it was
  the endpoint of at most $(i-1)+k-1$ paths for $t$ before this
  iteration; even if all the $k$ paths for $t$ in this iteration ended
  at $t'$, it is the endpoint of at most $i+2k-2$ paths for $t$ after
  the iteration. This gives us the desired property 1, completing the
  proof.
\end{proofof}

Theorem~\ref{thm:kconnAug} and Lemma~\ref{lem:elemConnectivity} have
applications to more general problems including the node-weighted
version of \sskconn\ \cite{ChuzhoyK08} and rent-or-buy and buy-at-bulk
network design \cite{ChekuriK08}. We omit discussion of these
applications in this version of the paper.

\section{Conclusions}

Having generalized the reduction step of \cite{hind} to handle local
element connectivity, we demonstrated applications of this stronger
Reduction Lemma to packing element (and edge) disjoint Steiner trees
and forests, and also to \sskconn. We believe that the Reduction Lemma
will find other applications in the future. We close with several open
questions:
\begin{itemize}
\item We believe that our bound on the number of element-disjoint
Steiner forests in a general graph can be improved from
$\Omega(k/(\log |T| \log m))$ to $\Omega(k/\log |T|)$.

\item Prove or disprove Conjecture~\ref{conj:minorfree}, on packing
  disjoint Steiner Forests in graphs excluding a fixed minor.

\item In a natural generalization of the Steiner Forest packing
  problem, each non-terminal/white vertex has a \emph{capacity}, and
  the goal is to pack forests subject to these capacity constraints.
  In general graphs, it is easy to reduce this problem to the
  uncapacitated/unit-capacity version (for example, by replacing a
  white vertex of capacity $c$ by a clique of size $c$), but this is
  not necessarily the case for restricted classes of graphs. In
  particular, it would be interesting to pack $\Omega(k)$ forests for
  the capacitated planar Steiner Forest problem.

\item The known hardness of approximation factor for \sskconn~is
  $\Omega(\log n)$ when $k$ is a polynomial function of $n$, the number
  of vertices \cite{KortsarzKL}. Can the current ratio of $O(k \log
  |T|)$ be improved? 
\end{itemize}

\paragraph{Acknowledgements:} We thank Anupam Gupta for several
discussions on single-sink $k$-vertex-connectivity and the potential
of connectivity-preserving reductions to give a simpler proof of
Theorem~\ref{thm:kconnAug}. We thank Sanjeev Khanna and Julia Chuzhoy
for sharing a preliminary version of \cite{ChuzhoyK08}.  We thank
Joseph Cheriyan for asking about planar packing of Steiner Trees which
inspired our work on that problem. We also thank Oleg Borodin, Dan
Cranston, Alexandr Kostochka and Doug West for pointers to structural
results on planar graphs.

\appendix

\section{Packing Element-Disjoint Trees and Forests}
\label{app:packing}

\subsubsection*{A Counterexample to the Random Coloring algorithm for
packing Steiner Forests.}\label{subsec:counterexample}

We first define a graph $H_k$, which we use subsequently. $H_k$ has
two black vertices $x$ and $y$, and $k$ white vertices, each incident
to both $x$ and $y$. (That is, there are $k$ disjoint paths of white
vertices from $x$ to $y$.) Given a graph $G$, we define the operation
of inserting $H_k$ along an edge $pq \in E(G)$ as follows: Add the
vertices and edges of $H_k$ to $G$, delete the edge $pq$, and add
edges from $p$ to $x$ and $q$ to $y$. (If we collapsed $H_k$ to a
single vertex, we would have subdivided the edge $pq$.) Figure 2 below
shows $H_4$ and the effect of inserting $H_4$ along an edge.

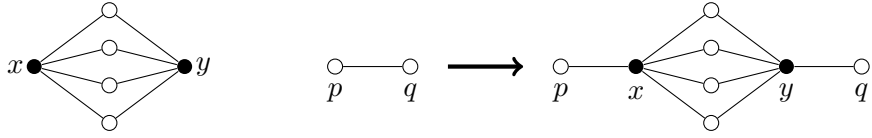
\begin{figure}[h]
  \begin{center}
    \begin{tikzpicture}

      \tikzstyle{dot}=[circle,inner sep=2pt,fill=black];
      \tikzstyle{elem}=[circle,draw,inner sep=2pt,fill=white];

      \begin{scope}
        \node (x) at (1,1) [dot] {}; \node at (0.75,1) {$x$};
        \node (y) at (3,1) [dot] {}; \node at (3.25,1) {$y$};
        
        \node (w1) at (2,1.75) [elem] {}; \node (w2) at (2,1.25) [elem] {};
        \node (w3) at (2,0.75) [elem] {}; \node (w4) at (2,0.25) [elem] {};

        \draw (x) -- (w1) -- (y) -- (w2) -- (x) -- (w3) -- (y) -- (w4)
        -- (x);
      \end{scope}

      \begin{scope} [xshift=8cm]

        \node (p) at (-3,1) [elem] {}; \node at (-3,0.65) {$p$};
        \node (q) at (-2,1) [elem] {}; \node at  (-2,0.65) {$q$};

        \draw (p) -- (q);
        \draw[->, ultra thick] (-1.5,1) -- (-0.5,1);

        \node (p1) at (0,1) [elem] {}; \node at (0,0.65) {$p$};
        \node (q1) at (4,1) [elem] {}; \node at (4,0.65) {$q$};
        
        \node (x) at (1,1) [dot] {}; \node at (1,0.65) {$x$};
        \node (y) at (3,1) [dot] {}; \node at (3,0.65) {$y$};
        
        \node (w1) at (2,1.75) [elem] {}; \node (w2) at (2,1.25) [elem] {};
        \node (w3) at (2,0.75) [elem] {}; \node (w4) at (2,0.25) [elem] {};

        \draw (x) -- (w1) -- (y) -- (w2) -- (x) -- (w3) -- (y) -- (w4)
        -- (x);

        \draw (p1) -- (x) (y) -- (q1);
        
      \end{scope}
    \end{tikzpicture}
    \caption{On the left, the graph $H_4$. On the right, inserting it along a single edge $pq$.}
  \end{center}
\end{figure}

We now describe the construction of our counterexample. We begin with
2 black vertices $s$ and $t$, and $k$ vertex-disjoint paths between
them, each of length $k+1$; there are no edges besides the ones just
described. Each of the $k^2$ vertices besides $s$ and $t$ is white. It
is obvious that $s$ and $t$ are $k$-element-connected in this
graph. Now, to form our final graph $G_k$, insert a copy of $H_k$
along each of the $k(k-1)$ edges between a pair of white vertices.
Fig. 3 below shows the construction of $G_3$.

\begin{figure}[h]
  \begin{center}
    \begin{tikzpicture}

      \tikzstyle{dot}=[circle,inner sep=2pt,fill=black];
      \tikzstyle{elem}=[circle,draw,inner sep=2pt,fill=white];

      \begin{scope}
        \node (s) at (0,1) [dot] {}; \node at (-0.25,1) {$s$};
        \node (t) at (4,1) [dot] {}; \node at (4.25,1) {$t$};
        
        \draw (s) -- (t) -- (3,2) -- (1,2) -- (s) -- (1,0) -- (3,0) -- (t);
        
        \foreach \x in {1,2,3} 
          \foreach \y in {0,1,2,}
          {
            \node at (\x,\y) [elem] {};
          }          
        
        \draw[->, ultra thick] (5,1) -- (7,1);        
      \end{scope}

      \begin{scope} [xshift=8cm]
        \node (s) at (0,1) [dot] {}; \node at (-0.25,1) {$s$};
        \node (t) at (6,1) [dot] {}; \node at (6.25,1) {$t$};
        
        \draw (s) -- (t) -- (5,3) -- (1,3) -- (s) -- (1,-1) -- (5,-1) -- (t);
        
        \foreach \x in {1,3,5} 
          \foreach \y in {-1,1,3}
          {
            \node at (\x,\y) [elem] {};
          }            
        \foreach \x in {2,4}
          \foreach \y in {-1,1,3}
          {
            \draw[fill=white] (\x,\y) ellipse (0.5cm and 0.9cm); \node at (\x,\y) {$H_3$};
          }
      \end{scope}
    \end{tikzpicture}
    \caption{ The construction of $G_3$.}
  \end{center}
\end{figure}
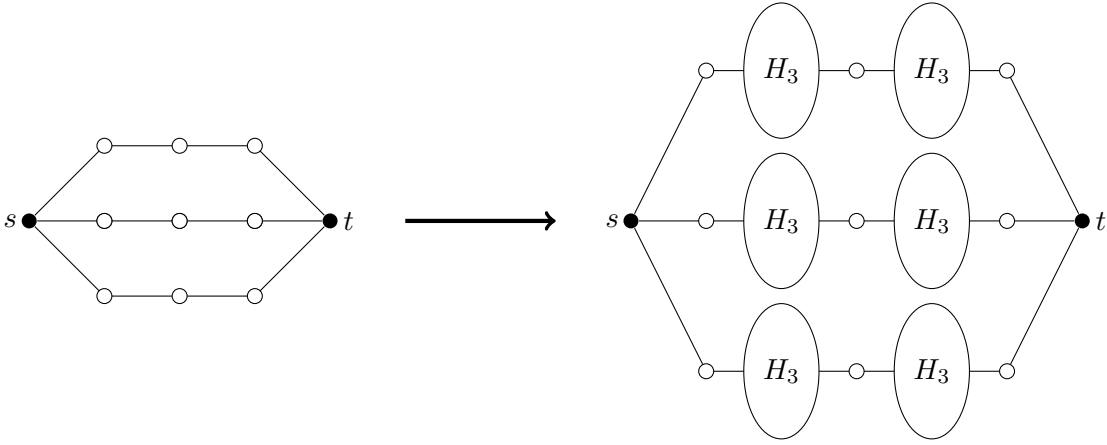

The following claims are immediate:
\begin{itemize}
\item The vertices $s$ and $t$ are $k$-element-connected in $G_k$.

\item For every copy of $H_k$, the vertices $x$ and $y$ are $k$-white connected in $G_k$.

\item The graph $G_k$ is bipartite, with the white vertices and the black vertices forming the two parts.
\end{itemize}

We use $G_k$ as an instance of the Steiner-forest packing problem; $s$
and $t$ form one group of terminals, and for each copy of $H_k$, the
vertices $x$ and $y$ of that copy form a group. From our claims above,
each group is $k$-element-connected.\\

If we use the algorithm of Cheriyan and Salavatipour, there are no
edges between white vertices to be deleted or contracted, so we move
directly to the coloring phase. If colors are assigned to the white
vertices randomly, it is easy to see that no color class is likely to
connect up $s$ and $t$. The probability that a white vertex is given
color $i$ is $\frac{c \log |T|}{k}$, for some constant $c$. The vertices
$s$ and $t$ can be connected iff the same color is assigned to all the
white vertices on one of the $k$ paths from $s$ to $t$ in the graph
formed from $G_k$ by contracting each $H_k$ to a single vertex. The
probability that \emph{every} vertex on such a path will receive the
same color is $\left(\frac{c \log |T|}{k} \right)^{k}$;
using the union bound over the $k$ paths gives us the desired result.

\subsection{Packing Trees in Planar Graphs}
\label{app:planarPacking}

\begin{lemma}\label{lem:discharging}
  Let $G(T \cup W, E)$ be a planar graph with minimum degree 3, in
  which $W$ is a stable set. There exists a vertex $t \in T$ of
  degree at most $10$, with at most $5$ neighbors in $T$. 
\end{lemma}
\begin{proof}
  Our proof uses the \emph{discharging} technique. Assume, for the
  sake of contradiction, that every vertex $t \in T$ has degree at
  least 11, or has at least 6 neighbors in $T$. By multiplying Euler's
  formula by 4, we observe that for a planar graph $G(V,E)$ with face
  set $F$,  $(2|E| - 4|V|) + (2|E| - 4|F|) = -8$. We rewrite this as
  $\sum_{v \in V} \left(d(v) - 4 \right) + \sum_{f \in F} \left( l(f)
    - 4 \right) = -8$, where $d(v)$ and $l(f)$ denote the degree of
  vertex $v$ and length of face $f$ respectively.

  Now, in our given graph $G$, assign $d(v) - 4$ units of
  \emph{charge} to each vertex $v \in T \cup W$, and assign $l(f) - 4$
  units of charge to each face $f$: Note that the net charge on the
  graph is negative. (It is equal to $-8$.) We describe rules for
  redistributing the charge through the graph such that after
  redistribution, if every vertex $t \in T$ has degree at least 11 or
  has at least 6 neighbors in $T$, the charge at each vertex and face
  will be non-negative. But no charge is added or removed (it is merely
  rearranged), and so we obtain a contradiction.

  \bigskip \noindent
  We use the following rules for distributing charge:
  \begin{enumerate}
    \item Every terminal $t \in T$ distributes $1/3$ unit of charge to
      each of its neighbors in $W$.
    \item Every terminal $t \in T$ distributes $1/2$ unit of charge to
      each triangular face $f$ it is incident to, unless the face
      contains 3 terminals. In this case, it distributes $1/3$ unit of
      charge to the face.
  \end{enumerate}

  We now observe that every vertex of $W$ and every face has
  non-negative charge. Each vertex $u \in W$ has degree at least $3$
  (the graph has minimum degree $3$), so its initial charge was at
  least $-1$. It did not give up any charge, and rule 1 implies that
  it received $1/3$ from each of its (at least $3$) neighbors, all of
  which are in $T$. Therefore, $u$ has non-negative charge after
  redistribution. If a face $f$ has length $4$ or more, it already had
  non-negative charge, and it did not give up any. If $f$ is a
  triangle, it starts with charge $-1$. It is incident to at least $2$
  terminals, since $W$ is a stable set; we argue that it gains $1$
  unit of charge, to end with charge $0$. From rule 2, if $f$ is
  incident to 2 terminals, it gains $1/2$ unit from each of them, and
  if it is adjacent to $3$ terminals, it gains $1/3$ unit from each of
  them. 

  \bigskip It remains only to argue that each terminal $t \in T$ has
  non-negative charge after redistribution. For ease of analysis, we
  describe a slightly modified version of the discharging in which
  each terminal loses at least as much charge as under the original
  rules, and show that each terminal has non-negative charge under the
  new discharging rules, listed below:
  \begin{enumerate}
  \item Every terminal $t$ gives $1/3$ unit of charge to \emph{every}
    neighbor.

  \item Every terminal $t \in T$ gives $1/3$ unit of charge to each
    adjacent triangle.

  \item Every terminal $t$ gets back $1/3$ unit of charge from each face
    $f$ such that both $t$'s neighbors on $f$ are black.
  \end{enumerate}
  
  We first prove that every terminal $t$ loses at least as much charge
  as under the original rules; see also Fig.~\ref{fig:discharging}.
  The terminal $t$ is now giving $1/3$ unit of charge to all its black
  neighbors, besides giving this charge to its white neighbors. It is
  giving less charge ($1/3$ instead of $1/2$) to some triangular
  neighbors, but every triangle is incident to a black vertex $t'$
  besides $t$; this neighbor of $t$ received an extra $1/3$ unit of
  charge from $t$, and it can give $1/6 = 1/2-1/3$ to each face
  incident to the edge $t-t'$. That is, the extra charge of $1/3$
  given by $t$ to $t'$ is enough to compensate for the fact that $t$
  may give $1/6$ units less charge to the two faces incident to
  $t-t'$. Finally, note that if both $t$'s neighbors on some face $f$
  are black, the original rules require $t$ to give only $1/3$ unit to
  $f$, which it also does under the new rules. However, it has given
  $1/3$ unit of charge to these two black neighbors, and they do not
  need to use this to compensate for $t$ giving too little charge to
  $f$; therefore, they may each return $1/6$ unit of charge to $t$.

  \begin{figure}[h]
    \begin{center}
      \begin{tikzpicture}[scale=1.25]

        \tikzstyle{dot}=[circle,inner sep=2pt,fill=black];
        \tikzstyle{elem}=[circle,draw,inner sep=2pt,fill=white];
        \tikzstyle{every node}=[font=\footnotesize];
  
        \begin{scope}
          \node (t) at (0,0) [dot] {}; \node (t1) at (0,1.5) [dot] {};
          \node (w1) at (-1.5,1) [elem] {}; \node (w2) at (1.5,1) [elem] {};

          \draw (t) -- (w1) -- (t1) -- (w2) -- (t) -- (t1);

          \draw[very thick, ->] (t) -- (-0.75,0.5); \node at (-0.75,0.15) {$\nicefrac{1}{3}$};
          \draw[very thick, ->] (t) -- (0.75,0.5); \node at (0.75,0.15) {$\nicefrac{1}{3}$};
          \draw[very thick, ->] (t) -- (-0.5,0.75); \node at (-0.8,0.8) {$\nicefrac{1}{2}$};
          \draw[very thick, ->] (t) -- (0.5,0.75); \node at (0.8,0.8) {$\nicefrac{1}{2}$};

          \node at (0,-0.5) {(a): Old Rules.};
        \end{scope}

        \begin{scope}[xshift=4cm]
          \node (t) at (0,0) [dot] {}; \node (t1) at (0,1.5) [dot] {};
          \node (w1) at (-1.5,1) [elem] {}; \node (w2) at (1.5,1) [elem] {};

          \draw (t) -- (w1) -- (t1) -- (w2) -- (t) -- (t1);

          \draw[very thick, ->] (t) -- (-0.75,0.5); \node at (-0.75,0.15) {$\nicefrac{1}{3}$};
          \draw[very thick, ->] (t) -- (0.75,0.5); \node at (0.75,0.15) {$\nicefrac{1}{3}$};
          \draw[very thick, ->] (t) -- (0,1.2); 
          \draw[very thick, ->] (0,1.25) -- (-0.45,0.8); \node at (-0.5,1.1) {$\nicefrac{1}{6}$};
          \draw[very thick, ->] (0,1.25) -- (0.45,0.8); \node at (0.5,1.1) {$\nicefrac{1}{6}$};
          \draw[very thick, ->] (t) -- (-0.5,0.75); \node at (-0.8,0.8) {$\nicefrac{1}{3}$};
          \draw[very thick, ->] (t) -- (0.5,0.75); \node at (0.8,0.8) {$\nicefrac{1}{3}$};

          \node at (0,-0.5) {(b): Equivalence of the rules};
        \end{scope}

        \begin{scope}[xshift=8cm]
          \node (t) at (0,0) [dot] {}; \node (t1) at (0,1.5) [dot] {};
          \node (w1) at (-1.5,1) [elem] {}; \node (w2) at (1.5,1) [elem] {};
          \draw (t) -- (w1) -- (t1) -- (w2) -- (t) -- (t1);

          \draw[very thick, ->] (t) -- (-0.75,0.5); \node at (-0.75,0.15) {$\nicefrac{1}{3}$};
          \draw[very thick, ->] (t) -- (0.75,0.5); \node at (0.75,0.15) {$\nicefrac{1}{3}$};
          \draw[very thick, ->] (t) -- (0,1.1); \node at (0.2,1.2) {$\nicefrac{1}{3}$};
          \draw[very thick, ->] (t) -- (-0.5,0.75); \node at (-0.8,0.8) {$\nicefrac{1}{3}$};
          \draw[very thick, ->] (t) -- (0.5,0.75); \node at (0.8,0.8) {$\nicefrac{1}{3}$};

          \node at (0,-0.5) {(c): New Rules.};
        \end{scope}

        \begin{scope}[xshift=0cm,yshift=-2.5cm]

          \node (t) at (0,0) [dot] {};
          \node (t1) at (-1,1) [dot] {}; \node (t2) at (1,1) [dot] {};
          \node (w1) at (-1.5,0) [elem] {}; \node (w2) at (1.5,0) [elem] {};

          \draw (t) -- (t1) -- (t2) -- (t);
          \draw (t1) -- (w1) -- (t) -- (w2) -- (t2);

          \draw[very thick,->] (t) -- (0,0.5); \node at (0,0.7) {$\nicefrac{1}{3}$};
          \draw[very thick,->] (t) -- (-0.625,0.25); \node at (-0.8,0.25) {$\nicefrac{1}{2}$};
          \draw[very thick,->] (t) -- (0.625,0.25); \node at (0.8,0.25) {$\nicefrac{1}{2}$};
          \draw[very thick,->] (t) -- (-0.75,0); \node at (-0.75,-0.25) {$\nicefrac{1}{3}$};
          \draw[very thick,->] (t) -- (0.75,0); \node at (0.75,-0.25) {$\nicefrac{1}{3}$};

          \node at (0,-0.7) {(d): Old Rules.};
        \end{scope}

        \begin{scope}[xshift=4cm,yshift=-2.5cm]

          \node (t) at (0,0) [dot] {};
          \node (t1) at (-1,1) [dot] {}; \node (t2) at (1,1) [dot] {};
          \node (w1) at (-1.5,0) [elem] {}; \node (w2) at (1.5,0) [elem] {};

          \draw (t) -- (t1) -- (t2) -- (t);
          \draw (t1) -- (w1) -- (t) -- (w2) -- (t2);

          \draw[very thick,->] (t) -- (-0.625,0.25); \node at (-0.8,0.2) {$\nicefrac{1}{3}$};
          \draw[very thick,->] (t) -- (0.625,0.25); \node at (0.8,0.2) {$\nicefrac{1}{3}$};
          \draw[very thick,->] (t) -- (-0.75,0); \node at (-0.75,-0.25) {$\nicefrac{1}{3}$};
          \draw[very thick,->] (t) -- (0.75,0); \node at (0.75,-0.25) {$\nicefrac{1}{3}$};
          \draw[very thick,->] (t) -- (-0.7,0.7); 
          \draw[very thick,->] (t) -- (0.7,0.7); 

          \draw[very thick,->] (-0.75,0.75) -- (0,0.75); \node at (-0.4,0.87) {$\nicefrac{1}{6}$};
          \draw[very thick,->] (-0.75,0.75) -- (-0.75,0.35); \node at (-0.95,0.55) {$\nicefrac{1}{6}$};
          \draw[very thick,->] (0.75,0.75) -- (0,0.75); \node at (0.4,0.87) {$\nicefrac{1}{6}$};
          \draw[very thick,->] (0.75,0.75) -- (0.75,0.35); \node at (0.95,0.55) {$\nicefrac{1}{6}$};

          \node at (0,-0.7) {(e): Equivalence of the rules};
        \end{scope}

        \begin{scope}[xshift=8cm,yshift=-2.5cm]

          \node (t) at (0,0) [dot] {};
          \node (t1) at (-1,1) [dot] {}; \node (t2) at (1,1) [dot] {};
          \node (w1) at (-1.5,0) [elem] {}; \node (w2) at (1.5,0) [elem] {};

          \draw (t) -- (t1) -- (t2) -- (t);
          \draw (t1) -- (w1) -- (t) -- (w2) -- (t2);

          \draw[very thick,->] (t) -- (-0.625,0.25); \node at (-0.8,0.25) {$\nicefrac{1}{3}$};
          \draw[very thick,->] (t) -- (0.625,0.25); \node at (0.8,0.25) {$\nicefrac{1}{3}$};
          \draw[very thick,->] (t) -- (-0.75,0); \node at (-0.75,-0.25) {$\nicefrac{1}{3}$};
          \draw[very thick,->] (t) -- (0.75,0); \node at (0.75,-0.25) {$\nicefrac{1}{3}$};
          \draw[very thick,->] (t) -- (-0.75,0.75); \node at (-0.45,0.75) {$\nicefrac{1}{3}$};
          \draw[very thick,->] (t) -- (0.75,0.75); \node at (0.45,0.75) {$\nicefrac{1}{3}$};

          \node at (0,-0.7) {(f): New Rules.};
        \end{scope}

      \end{tikzpicture}
    \end{center}
    \captionsetup{justification=centerfirst,font=small}
    \caption{Terminals lose at least as much charge under the new
      rules. \newline
      Part (a) shows the charge given away by a
      terminal under the original rules, while part (c) shows the
      charge given away under the new rules; the triangles now receive
      less charge.  Part (b) shows that the extra 1/3 unit of charge
      given to the black neighbor under the new rules can be split
      equally among the two triangles, which has the same effect as
      giving 1/2 unit to the triangles.
      Similarly, part (d) shows the charge given away by a terminal
      under the original rules, while part (f) shows the charge under
      the new rule 3: The central triangular face receives 1/3 unit of
      charge, but also returns 1/3 charge to the terminal as both its
      neighbors on this face are black. Part (e) shows that the extra
      1/3 unit of charge given to each black neighbor under the new
      rules can be split among the triangles, so the effect is the
      same as giving 1/3 unit of charge to the central face, and 1/2
      to each of the other faces.
    } \label{fig:discharging}
  \end{figure}
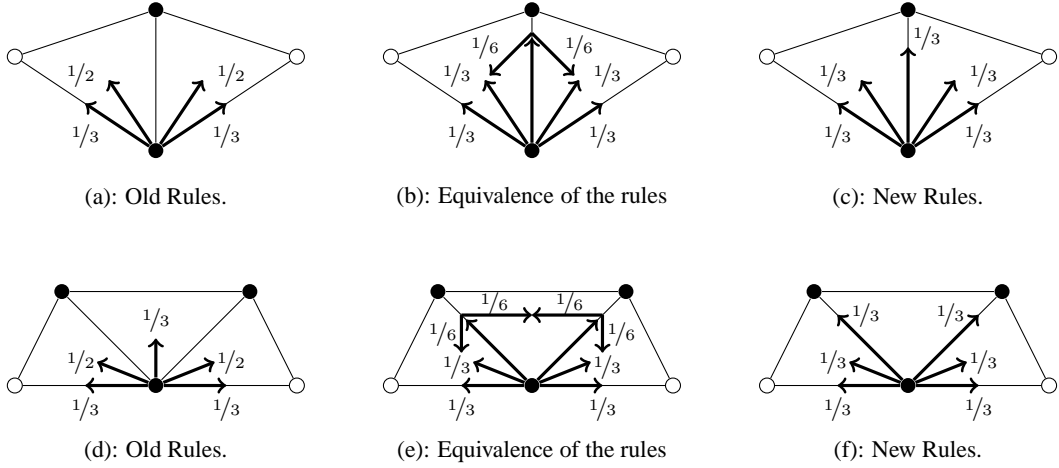

  We now argue that every terminal has non-negative charge under the
  new rules. Let $t \in T$ have degree $d$; we consider three cases:
  \begin{enumerate}
  \item If $d \ge 12$, $t$ gives away $1/3$ to each of its $d$
    neighbors and $d$ incident faces, so the total charge it gives
    away is $2d/3$. (It may also receive some charge, but we ignore
    this.) Therefore, the net charge on $t$ is $(d - 4) - 2d/3 = (d/3)
    - 4$; as $d \ge 12$, this cannot be negative.

  \item If $d = 11$, we count the number of triangles incident to
    $t$. If there are 10 or fewer, $t$ gives away $1/3$ unit of charge
    to each of its 11 neighbors, and at most $10/3$ to its adjacent
    triangles, so the net charge on $t$ is at least $(11 - 4) - 11/3 -
    10/3 = 0$. If $t$ is incident to 11 triangles, it must be adjacent
    to at least 6 black vertices, as each triangle incident to $t$
    must be adjacent to a black neighbor of $t$, and no more than 2
    triangles incident to $t$ can share a neighbor of $t$. Since $t$
    has degree 11 and at least 6 black neighbors, some pair of black
    neighbors of $t$ are on a common face, and $t$ must receive $1/3$
    unit of charge from this face. It follows that the charge on $t$
    is at least $(11 - 4) - 11/3 - 11/3 + 1/3 = 0$. 

  \item If $d \le 10$, $t$ has at least 6 black neighbors by
    hypothesis. It has at most $d - 6$ white neighbors, so there are
    at least $6 - (d - 6) = 12 - d$ faces $f$ such that both $t$'s
    neighbors on $f$ are black. (Delete the white neighbors; there are
    at least $6$ faces incident to $t$ on which both its neighbors are
    black. When each white vertex is added back, it can only decrease
    the number of such faces by 1.) The terminal $t$ gives away $1/3$
    unit of charge to each of its $d$ neighbors and at most $d$
    incident triangles, and receives $1/3$ unit of charge from each
    face on which both its neighbors are black. Therefore, the net
    charge on $t$ is at least $(d - 4) - 2d/3 + (12 - d)/3 = 0$.
  \end{enumerate}
  \vspace{-0.35in}
\end{proof}

\bigskip
\begin{proofof}{Lemma~\ref{lem:parallelEdges}}
  Our argument is very similar to that of the proof in
  Section~\ref{subsec:planarPacking} that there are two terminals with
    at least $\ceil{k/10}$ edges between them, except that here we use
    Lemma~\ref{lem:discharging} instead of Theorem~\ref{thm:Borodin}.

    Let $G$ be the planar multigraph of the reduced instance; every
    terminal has degree at least $k$ in $G$. Construct a planar graph
    $G'$ from $G$ by keeping a single copy of each edge; from
    Lemma~\ref{lem:discharging} above, some terminal $t$ has degree at
    most 10, and at most 5 black neighbors. Let $w$ denote the number
    of white neighbors of $t$, and $b$ the number of black neighbors.
    Since each white vertex is incident to only a single copy of each
    edge in $G$, there must be at least $\ceil{(k-w)/b}$ copies in
    $G$ of some edge between $t$ and a black neighbor. But $b \le 5$
    and $b+w \le 10$; it is easy to verify since $k \ge 10$, the
    smallest possible value of $\ceil{(k-w)/b}$ is $\ceil{(k-5)/5} =
    \ceil{k/5} - 1$.
\end{proofof}

\subsection{An Algorithm for Packing Steiner Forests in Planar and
  Bounded-genus Graphs}
\label{subsec:forestPacking}

For the Planar Steiner Forest Packing problem, we use an algorithm
very similar to that for packing Steiner Trees in
Section~\ref{subsec:planarPacking}. Now, as input, we are given sets
$T_1, \ldots T_m$ of terminals that are each internally $k$-connected,
but some $T_i$ and $T_j$ may be poorly connected. Precisely as before,
as long as each $T_i$ contains at least 2 terminals,
Lemma~\ref{lem:parallelEdges} is true, so we can contract some pair of
terminals $t_1, t_2$ that have $\ceil{k/5}-1$ parallel edges between
them. Note that if $t_1, t_2$ are in the same $T_i$, after
contraction, we have an instance in which $T_i$ contains fewer
terminals, and we can apply induction. If $t_1, t_2$ are in different
sets $T_i, T_j$, then after contracting, all terminals in $T_i$ and
$T_j$ are pairwise $k$-connected, so we can merge these two groups
into a single set.

In proving the crucial Lemma~\ref{lem:parallelEdges}, we argued that
in the multigraph $G$ of the reduced instance, every terminal has
degree at least $k$ (since it is $k$-element-connected to other
terminals), and in the graph $G'$ in which we keep only a single copy
of each edge, some terminal has degree at most 10; therefore, there
are $\ceil{k/10}$ copies of some edge. However, in the Steiner Forest
problem, some $T_i$ may contain only a single terminal $t$ (after
several contraction steps). The terminal $t$ may be poorly connected
to the remaining terminals; therefore, it may have degree less than
$k$ in the multigraph $G$. If $t$ is the unique low-degree terminal in
$G'$, we may not be able to find a pair of terminals with a large
number of edges between them. As a concrete example, consider the
graph $G_k$ defined at the beginning of this appendix. (See also
Fig. 3, and note that $G_k$ is planar.) We have one terminal set $T_1
= \{s, t\}$, and other sets $T_i$ containing the two terminals of each
copy of $H_k$. After several contraction steps, each copy of $H_k$ may
have been contracted together to form a single terminal; each such
terminal is only 2-connected to the rest of the graph. In the reduced
instance, there is only a single copy of each edge, and
Lemma~\ref{lem:parallelEdges} does not hold. 

We solve this problem by eliminating a set $T_i$ when it has only a
single terminal; at this point, we can apply induction and proceed. We
formalize this intuition in the following lemma:

\begin{lemma}
  Let $G(V, E)$ with a given $T \subseteq V$ be a planar graph, and $t
  \in T$ be an arbitrary terminal of degree $d$. Let $G'$ be the graph
  constructed from $G$ by deleting $t$, and inserting a $d \times d$
  grid of white vertices, with the edges incident to $t$ in $G$ made
  incident to distinct vertices on one side of the new grid in
  $G'$. Then:
  \begin{enumerate}
    \item $G'$ is planar.
    \item For every pair $u, v$ of terminals in $G'$,
      $\elconn_{G'}(u, v) = \elconn_{G}(u, v)$.
    \item Any set of element-disjoint subgraphs of $G'$ corresponds to a
      set of element-disjoint subgraphs of $G$.
  \end{enumerate}
\end{lemma}

\begin{proofsketch}
  See Figure~\ref{fig:grid} showing this operation; it is easy to
  observe that given a planar embedding of $G$, one can construct a
  planar embedding of $G'$. It is also clear that a set of
  element-disjoint subgraphs in $G'$ correspond to such a set in $G$;
  every subgraph that uses a vertex of the grid can contain the
  terminal $t$.

  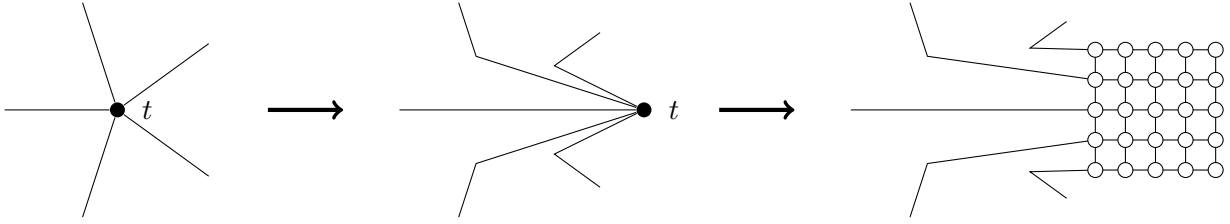
\begin{figure} 
    \begin{center}
      \begin{tikzpicture}
        
        \tikzstyle{dot}=[circle,inner sep=2pt,fill=black];
        \tikzstyle{elem}=[circle,draw,inner sep=2pt,fill=white];
        
        \begin{scope}[xshift=-5cm] 

          \node (t) at (0,0) [dot] {}; \node at (0.4,0) {$t$};
          
          \draw (36:1.5cm) -- (t) -- (-36:1.5cm); \draw (108:1.5cm) -- (t) --
          (-108:1.5cm); \draw (t) -- (180:1.5cm);
          
          \draw[->,ultra thick] (2,0) -- (3,0);

        \end{scope}
        
        \begin{scope} 

          \node (t) at (2,0) [dot] {}; \node at (2.4,0) {$t$};
          
          \draw (36:1.75cm) -- (36:1cm) -- (t);
          \draw (-36:1.75cm) -- (-36:1cm) -- (t);
          \draw (-108:1.5cm) -- (-108:0.75cm) -- (t);
          \draw (108:1.5cm) -- (108:0.75cm) -- (t);
          \draw (180:1.25cm) -- (180:1cm) -- (t);
          
          \draw [->, ultra thick] (3,0) -- (4,0);
        \end{scope}

        \begin{scope}[xshift=6cm] 

          \draw [step=0.4cm] (1.9,-0.8) grid (3.6,0.8);

          \draw (36:2cm) -- (36:1.4cm) -- (2,0.8);
          \draw (-36:2cm) -- (-36:1.4cm) -- (2,-0.8);
          \draw (-108:1.5cm) -- (-108:0.75cm) -- (2,-0.4);
          \draw (108:1.5cm) -- (108:0.75cm) -- (2,0.4);
          \draw (180:1.25cm) -- (180:1cm) -- (2,0);

          \foreach \y in {0.8,0.4,0,-0.4,-0.8} 
          {
            \foreach \x in {2,2.4,2.8,3.2,3.6} 
            {
              \node at (\x,\y) [elem] {};
            }
          }
        \end{scope}

      \end{tikzpicture}
    \end{center}
    \caption{Replacing a terminal by a grid of white vertices
      preserves planarity and element-connectivity.} \label{fig:grid}
  \end{figure}

  It remains only to argue that the element-connectivity of every
  other pair of terminals is preserved. Let $u,v$ be an arbitrary pair
  of terminals; we show that their element-connectivity in $G'$ is at
  least their connectivity $\elconn(u,v)$ in $G$. Fix a set of
  $\elconn(u,v)$ paths in $G$ from $u$ to $v$; let $\script{P}$ be the
  paths that use the terminal $t$, and let $\ell = |\script{P}|$. We
  locally modify these $\ell$ paths in $\script{P}$ by routing them through
  the grid, so we obtain $\elconn(u,v)$ element-disjoint paths in
  $G'$.

  Let $\script{P}_u$ denote the set of prefixes from $u$ to $t$ of the
  $\ell$ paths in $\script{P}$, and let $\script{P}_v$ denote the
  suffixes from $t$ to $v$ of these paths. Let $H$ denote the $d
  \times d$ grid that replaces $t$ in $G'$; we use $\script{P}'_u$ and
  $\script{P}'_v$ to denote the corresponding paths in $G'$ from $u$
  to vertices of $H$, and from vertices in $H$ to $v$ respectively.
  Let $\script{I}$ and $\script{O}$ denote the vertices of $H$
  incident to paths in $\script{P}'_u$ and $\script{P}'_v$.  It is not
  difficult to see that there are a set of disjoint paths in the grid
  $H$ connecting the $\ell$ distinct vertices in $\script{I}$ to those
  in $\script{O}$; using the paths of $\script{P}'_u$, together with
  the paths through $H$ and the paths of $\script{P}'_v$ gives us a
  set of disjoint paths in $G'$ from $u$ to $v$.
\end{proofsketch}

\medskip
\paragraph{A Counterexample to the existence of 2 terminals with $\Omega(k)$
  ``Parallel edges'' between them:}

Recall that in the case of planar graphs (or graphs of bounded genus),
we argued that there must be two terminals $t_1, t_2$ with $\Omega(k)$
``parallel edges'' between them. (That is, there are $\Omega(k)$
degree-2 white vertices adjacent to $t_1$ and $t_2$.)  This is not
necessarily the case even in graphs of treewidth 3:
The graph $K_{3,k}$, the complete bipartite graph with 3 vertices on
one side and $k$ on the other, has treewidth 3. If the three vertices
on one side are the terminal set $T$ and the $k$ vertices of the other
side are non-terminals, it is easy to see that $\elconn(T) = k$, but
every white vertex has degree 3.

In this example, there are only 3 terminals, so the tree-packing
algorithm of Cheriyan and Salavatipour \cite{cs} would allow us to
find $\Omega(k/\log |T|) = \Omega(k)$ trees connecting them. Adding
more terminals incident to all the white vertices would raise the
treewidth, so this example does not immediately give us a
low-treewidth graph with a large terminal set such that there are few
parallel edges between any pair of terminals. However, we can easily
extend the example by defining a graph $G_m$ as follows: Let $T_1,
T_2, \ldots T_m$ be sets of 2 terminals each, let $W_1, W_2, \ldots
W_{m-1}$ each be sets of $k$ white vertices, and let all the vertices
in each $W_i$ be adjacent to both terminals in $T_i$ and both
terminals in $T_{i+1}$. (See Fig.~\ref{fig:treewidthExample} below.)
The graph $G_m$ has $2m$ terminals, $T = \bigcup_i T_i$ is
$k$-element-connected, and it is easy to verify that $G_m$ has
treewidth 4. However, every white vertex has degree 4, so there are no
``parallel edges'' between terminals. (One can modify this example to
construct a counterexample graph $G_m$ with treewidth 3 by removing one
terminal from each alternate $T_i$.)

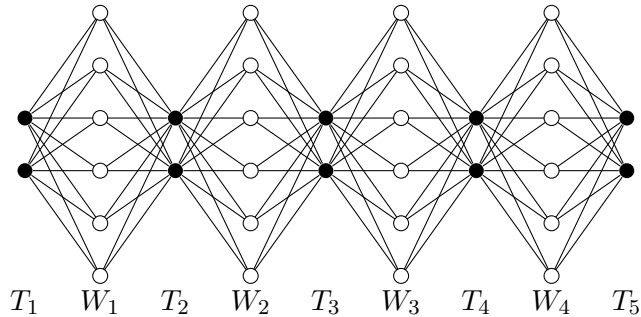
\begin{figure}[h]
  \begin{center}
    \begin{tikzpicture}[yscale=0.7]

      \tikzstyle{dot}=[circle,inner sep=2pt,fill=black];
      \tikzstyle{elem}=[circle,draw,inner sep=2pt,fill=white];
      
      \begin{scope}[xshift=0cm] 
          
        \node (t11) at (0,0.5) [dot] {}; \node (t12) at (0,-0.5) [dot] {};
        \node at (0,-3) {$T_1$};

        \foreach \y in {2.5,1.5,0.5,-0.5,-1.5,-2.5}
        {
          \draw (t11) -- (1,\y) -- (t12);
        }

        \node at (1,-3) {$W_1$};
      \end{scope}
      
      \begin{scope}[xshift=2cm] 
          
        \node (t21) at (0,0.5) [dot] {}; \node (t22) at (0,-0.5) [dot] {};
        \node at (0,-3) {$T_2$};

        \foreach \y in {2.5,1.5,0.5,-0.5,-1.5,-2.5}
        {
          \draw (t21) -- (1,\y) -- (t22) -- (-1,\y) -- (t21);
        }

        \node at (1,-3) {$W_2$};
      \end{scope}

      \begin{scope}[xshift=4cm] 
          
        \node (t31) at (0,0.5) [dot] {}; \node (t32) at (0,-0.5) [dot] {};
        \node at (0,-3) {$T_3$};

        \foreach \y in {2.5,1.5,0.5,-0.5,-1.5,-2.5}
        {
          \draw (t31) -- (1,\y) -- (t32) -- (-1,\y) -- (t31);
        }

        \node at (1,-3) {$W_3$};
      \end{scope}

      \begin{scope}[xshift=6cm] 
          
        \node (t41) at (0,0.5) [dot] {}; \node (t42) at (0,-0.5) [dot] {};
        \node at (0,-3) {$T_4$};

        \foreach \y in {2.5,1.5,0.5,-0.5,-1.5,-2.5}
        {
          \draw (t41) -- (1,\y) -- (t42) -- (-1,\y) -- (t41);
        }

        \node at (1,-3) {$W_4$};

        \node (t51) at (2,0.5) [dot] {}; \node (t52) at (2,-0.5) [dot] {};
        \node at (2,-3) {$T_5$};
        
        \foreach \y in {2.5,1.5,0.5,-0.5,-1.5,-2.5}
        {
          \draw (t51) -- (1,\y) -- (t52);
        }
      \end{scope}

      \foreach \x in {1,3,5,7}
      {
        \foreach \y in {2.5,1.5,0.5,-0.5,-1.5,-2.5}
        {
          \node at (\x, \y) [elem] {};
        }
      }

    \end{tikzpicture}
  \end{center}
  \vspace{-0.3in}
  \caption{A graph of treewidth 4 with many terminals, but no
    ``parallel edges''.} \label{fig:treewidthExample}
\end{figure}

\end{document}